\documentclass[aip,graphicx,reprint]{revtex4-1}

\draft 

\usepackage{amssymb}

\usepackage{amsmath}

\newcommand{\md}{\mathrm{d}}

\DeclareMathOperator*{\argmin}{arg\!\min}

\usepackage{empheq}

\DeclareSymbolFont{bbold}{U}{bbold}{m}{n}
\DeclareSymbolFontAlphabet{\mathbbold}{bbold}
\DeclareMathAlphabet{\mathpzc}{OT1}{pzc}{m}{it}

\newcommand{\angstrom}{{\mbox{\normalfont\AA}}}

\newcommand{\fd}{\mathbb{I}}

\newcommand{\bloss}{\underline{L}}

\newcommand{\oh}{\frac{1}{2}}

\newcommand{\pos}{\mathbf{R}}
\newcommand{\mom}{\mathbf{P}}
\newcommand{\fgpos}{\mathbf{r}}
\newcommand{\fgmom}{\mathbf{p}}
\newcommand{\prepos}{\mathpzc{r}}
\newcommand{\premom}{\mathpzc{p}}
\newcommand{\premass}{\mathpzc{m}}

\newcommand{\map}{\mathpzc{M}}
\newcommand{\rmap}{{\map_\fgpos}}
\newcommand{\pmap}{{\map_\fgmom}}
\newcommand{\vmap}{\mathpzc{G}}
\newcommand{\rvmap}{{\vmap_\prepos}}
\newcommand{\pvmap}{{\vmap_\premom}}
\newcommand{\Qual}{\mathcal{F}}

\newcommand{\cgParam}{{\boldsymbol{\theta}}}

\newcommand{\fgnp}{n}
\newcommand{\cgnp}{N}
\newcommand{\prenp}{\nu}

\newcommand{\preFeatDom}{{\mathcal{X}^\text{pre}}}
\newcommand{\preFeatDomPos}{{\mathcal{X}_\prepos^\text{pre}}}
\newcommand{\preFeatDomMom}{{\mathcal{X}_\premom^\text{pre}}}

\newcommand{\prePot}{{U^\text{pre}_\text{mod}}}
\newcommand{\prePotParam}{{U^\text{pre}_{\text{mod},\cgParam}}}

\newcommand{\preDensity}{{p^\text{pre}_\text{mod}}}
\newcommand{\preDensityParam}{{p^\text{pre}_{\text{mod},\cgParam}}}
\newcommand{\preDensityPos}{{p^\text{pre}_{\text{mod},\prepos}}}
\newcommand{\preDensityMom}{{p^\text{pre}_{\text{mod},\premom}}}

\newcommand{\dom}{\mathcal{X}}

\newcommand{\cgPot}{{U_\text{mod}}}

\newcommand{\cgDensity}{{p_\text{mod}}}
\newcommand{\cgDensityParam}{{p_{\text{mod},\cgParam}}}
\newcommand{\cgDensityPos}{{p_{\text{mod},\pos}}}
\newcommand{\cgDensityMom}{{p_{\text{mod},\mom}}}
\newcommand{\cgDensityPosParam}{{p_{\text{mod},\pos,\cgParam}}}

\newcommand{\cgPFPos}{{Z_{\text{mod},\pos}}}
\newcommand{\cgPFMom}{{Z_{\text{mod},\mom}}}

\newcommand{\refPot}{{U_\text{ref}}}

\newcommand{\refDensity}{{p_\text{ref}}}
\newcommand{\refDensityPos}{{p_{\text{ref},\pos}}}
\newcommand{\refDensityMom}{{p_{\text{ref},\mom}}}

\newcommand{\refFeatDom}{{\mathcal{X}^\text{FG}}}
\newcommand{\refFeatDomPos}{{\mathcal{X}^\text{FG}_\fgpos}}
\newcommand{\refFeatDomMom}{{\mathcal{X}^\text{FG}_\fgmom}}

\newcommand{\refPotMicro}{{U^\text{FG}_\text{ref}}}

\newcommand{\refDensityMicro}{{p^\text{FG}_\text{ref}}}
\newcommand{\refDensityMicroPos}{{p^\text{FG}_{\text{ref},\fgpos}}}
\newcommand{\refDensityMicroMom}{{p^\text{FG}_{\text{ref},\fgmom}}}

\newcommand{\refPFMicroPos}{{Z^\text{FG}_{\text{ref},\fgpos}}}
\newcommand{\refPFMicroMom}{{Z^\text{FG}_{\text{ref},\fgmom}}}

\newcommand{\varParam}{\boldsymbol{\psi}}
\newcommand{\fpairspace}{\mathcal{Q}}

\usepackage{amsthm}
\theoremstyle{definition}

\theoremstyle{remark}

\theoremstyle{theorem}

\begin{document}


\title{Adversarial-Residual-Coarse-Graining: Applying machine
learning theory to systematic molecular coarse-graining} 



\author{Aleksander E. P. Durumeric}
\email[]{aleksander@uchicago.edu}
\affiliation{Department of Chemistry, James Franck Institute,
Institute for Biophysical Dynamics, and Computation
Institute, The University of Chicago, Illinois 60637, USA}

\author{Gregory A. Voth}
\email[]{gavoth@uchicago.edu}
\affiliation{Department of Chemistry, James Franck Institute,
Institute for Biophysical Dynamics, and Computation
Institute, The University of Chicago, Illinois 60637, USA}

\date{\today}

\begin{abstract}
    We utilize connections between molecular coarse-graining approaches and
    implicit generative models in machine learning to describe a
    new framework for systematic molecular coarse-graining (CG). Focus is placed on the
    formalism encompassing generative adversarial networks.  The resulting
    method enables a variety of model parameterization strategies, some of which
    show similarity to previous CG methods.  We demonstrate that the
    resulting framework can rigorously parameterize CG models containing CG
    sites with no prescribed connection to the reference atomistic system
    (termed virtual sites); however, this advantage is offset by the lack of
    a closed-form expression for the CG Hamiltonian at the resolution obtained after
    integration over the
    virtual CG sites.  Computational examples are provided for cases in which
    these methods ideally return identical parameters as
    Relative Entropy Minimization
    (REM) CG but where traditional REM CG optimization equations are not 
    applicable.
\end{abstract}

\maketitle

\section{Introduction}

    Classical atomistic molecular dynamics (MD) simulation has provided
    significant insight into many biological and materials
    processes.\cite{karplus2002molecular,marx2009ab,
    dror2012biomolecular,karplus2014significance} However, its efficacy is
    often restricted by its computational cost: for example, routine atomic
    resolution studies of biomolecular systems are currently limited to
    microsecond simulations of millions of atoms.  Phenomena that cannot be
    characterized in this regime often require investigation using modified
    computational approaches. Coarse-grained (CG) molecular dynamics can be
    effective for studying systems where the motions of nearby atoms are highly
    interdependent.\cite{voth2008coarse,saunders2013coarse,noid2013perspective,
    brini2013systematic,baaden2013coarse}
    By simulating at the resolution of CG sites or ``beads'', 
    each associated with multiple
    correlated atoms, biomolecular processes at the second timescale and beyond
    can be accurately probed.  High-fidelity CGMD models often depend on 
    flexible parameterizations;
    as a result, the design of systematic
    parameterization strategies is an active area of study (e.g., methods
    and applications in references
    \citenum{stillinger1970effective,lyubartsev1995calculation,reith2003deriving,
    izvekov2005multiscale,
    noid2007multiscale,noid2008multiscale,noid2008multiscale2,shell2008relative,
    mullinax2009generalized,karimi2011ibisco,
    carmichael2012new,dama2013theory,rudzinski2015bottom,lyubartsev2015systematic,
    vlcek2015rigorous,de2016c,sanyal2016coarse,dunn2016bottom,
    lemke2017neural,john2017many,wagner2017extending,tsourtis2017parameterization,
    zhang2018deepcg}).

    The CGMD models considered in this article are similar to their atomistic
    counterparts. They are composed of point-mass CG beads, a corresponding CG
    force-field, and a simulation protocol that produces Boltzmann statistics in
    the long-time limit. We restrict the bulk of our study to the parameterization
    of the CG effective force-field.  Here, and in the remainder of the 
    article, we refer to
    these models simply as CG models.  We only consider the static
    equilibrium properties of these models, and not their
    dynamics. There are two nonexclusive
    classes of parameterization strategies for CG models of interest to this
    article:
        \emph{top-down} and
    \emph{bottom-up} approaches.\cite{voth2008coarse,noid2013perspective,
    saunders2013coarse}  Top-down approaches aim to parameterize CG
    models to recapitulate specific macroscopic properties, such as pressure and
    partition coefficients,\footnote{We note that recent
    work\cite{wagner2016representability} has reinforced how
    macroscopic observable matching has unexpected challenges when trying to
    establish a firm microscopic connection to atomistic systems if care is not
    taken in choosing which observable form to optimize with.} while bottom-up
    methods attempt to parameterize CG models to reproduce the multidimensional
    distribution given by explicitly mapping each atomistic configuration
    (produced by a suitable reference simulation) to a specific CG
    configuration.\cite{izvekov2005multiscale,
    noid2007multiscale,shell2008relative,noid2008multiscale,noid2008multiscale2}  
    The distribution of this mapped system is produced
    via a Boltzmann distribution with respect to an effective CG Hamiltonian
    referred to as the many-body Potential of Mean Force (PMF).

    Certain scientific inferences could be informally drawn from the fit CG
    force-field itself, assuming that the force-field is constrained to
    intuitive low dimensional contributions (e.g., pairwise
    forces, such as in ref \citenum{fan2012coarse}). For example,
    one could attempt to infer the effect of an amino acid mutation on protein
    behavior by considering how the approximated PMF differs when fit on
    reference wild type and mutant proteins simulations, similar to the analysis
    of low dimensional free energy surfaces. However,
    the primary use of CG models is typically based on their ability to generate
    CG configurations of a system of interest using their approximate
    force-field.
    \cite{srivastava2012hybrid,cao2012solvent,dunn2016bottom,carmichael2012new}
    The computational similarity of CG models with their atomistic counterparts
    allows CG models to be simulated using the same high performance software
    packages as those used in atomistic
    simulation.\cite{plimpton1995fast,abraham2015gromacs,
    brooks1983charmm,anderson2008general,case2005amber,phillips2005scalable,
    bowers2006scalable} As a result, the computational profile of CG models is
    often controlled by the same dominating factor as atomistic models: the
    calculation of the force-field at each timestep.\cite{plimpton1995fast,
    plimpton2012computational,markidis2015solving} This cost provides additional
    motivation for specific low dimensional force-field contributions.  However,
    there is no guarantee that a force-field characterized solely by traditional
    bonded and pairwise nonbonded terms either
    describes the true PMF of the CG variables or
    can accurately reproduce all observables of interest to the
    parameterization.\cite{voth2008coarse,
    noid2013perspective,saunders2013coarse} In the case of bottom-up methods,
    while typical approaches will produce the PMF in the infinite sampling limit
    when they are capable of representing any CG force-field, in practice each method
    creates a characteristic approximation (e.g.,
    reproducing two-body at the expense of higher order correlations).

    The compromises invoked by various bottom-up CG methods in
    realistic applications are critical to the utility of the resulting models.
    Certain methods focus on reproducing correlations dual to the potential form
    used;\cite{rudzinski2011coarse,lyubartsev1995calculation,chaimovich2011coarse,
    reith2003deriving} 
    for example, when using a
    pairwise nonbonded potential these methods recapitulate the radial
    distribution function of the target system. Other specific methods are
    characterized by attempting to reproduce both these dual correlations along
    with certain higher order correlations intrinsically connected to the CG
    potential.\cite{noid2007multiscale,noid2008multiscale,mullinax2009generalized,
    rudzinski2011coarse,izvekov2005multiscale,mullinax2009generalized,
    noid2007multiscale} The nature of
    the distributions approximated suggests three natural approaches for
    improving an inaccurate model: improve the CG force-field basis used, modify
    the CG representation, or select a different procedure to generate the CG
    force-field.  The first two options are often a central part of the design
    of a systematic CG model; however, realistic systems, such as proteins, may
    not be well described by correlations that are typically connected to 
    computationally
    efficient CG potentials coupled with appropriate CG
    representations.\cite{noid2013perspective}  More generally, the specific
    correlations critical to a reasonably accurate CG model may depend on the
    study at hand, and may be representable by simple force-fields---but only
    at the expense of other correlations connected to that
    potential form as dictated by a particular method. 
    As a result, the diversity of possible applications
    motivates the creation of additional strategies for bottom-up CG modeling,
    each of which has different biases in the approximations it produces.

    The task of generating examples (such as images) similar to a known
    empirical sample is of significant interest to the Machine Learning (ML)
    community.\cite{goodfellow2016nips,doersch2016tutorial,alain2016gsns,
    mohamed2016learning}  The creation of an artificial process that can produce
    realistic samples often entails encoding an understanding of the true
    mechanism underlying the real world distribution; internal representations
    of an accurately parameterized generative model, such as neural network
    parameters, can be transferred for use in secondary tasks such as
    classification\cite{2015arXiv151106434R} or image
    retrieval.\cite{2016arXiv160702748C}  The artificial samples produced by the
    models themselves have additionally shown value by providing novel
    molecular targets for synthesis
    \cite{sanchez2017optimizing,2017arXiv171207449E} or as labeled images for
    training in classification or
    regression.\cite{shafaei2016play,wood2016learning}  A substantial number of
    these complex applications utilize implicit generative models.
    \cite{mohamed2016learning,goodfellow2014generative,salakhutdinov2010efficient,
    2013arXiv1312.6114K}  Implicit generative models, such as Generative
    Adversarial Networks (GANs),\cite{goodfellow2014generative} are
    characterized by their lack of an explicit probability distribution, or
    an associated free energy, at the resolution they produce
    examples.\cite{mohamed2016learning}  For example, a GAN may be trained to
    generate pictures containing human faces.\cite{goodfellow2014generative}
    Each picture that could be generated has a parameterization specific
    probability of being a reasonable picture of a human face
    (admittedly, this probability is often very close to one or zero); however, the
    GAN itself does not have explicit knowledge of this probability.  Instead,
    the GAN is simply characterized as a procedure that transforms random
    numbers from a simple noise distribution to images that
    follow the probability distribution of plausible images. The methods used to
    parameterize (i.e.,
    train) GANs therefore focus on the ability to critique a model distribution
    against reference samples without knowledge of the probability density
    function characterizing the model.  This is in strong contrast to typical
    molecular simulation, \cite{karplus2002molecular,frenkel2002understanding,
    allen2017computer}
    which traditionally requires a known free energy surface to produce samples
    through molecular dynamics or Markov Chain Monte Carlo---and whose
    systematic parameterization techniques often naturally explicitly involve
    evaluation of the corresponding model free energy surface.
    \cite{stillinger1970effective,lyubartsev1995calculation,izvekov2005multiscale,
    noid2007multiscale,noid2008multiscale,noid2008multiscale2,shell2008relative,
    mullinax2009generalized,karimi2011ibisco,
    carmichael2012new,dama2013theory,rudzinski2015bottom,lyubartsev2015systematic,
    vlcek2015rigorous,de2016c,sanyal2016coarse,dunn2016bottom,
    lemke2017neural,john2017many,wagner2017extending,tsourtis2017parameterization,
    zhang2018deepcg}
    However, both methods are focused on accurately
    producing samples, or configurations, as their primary goal.  
    
    This article focuses on making this intuitive connection between GANs and
    molecular models explicit, allowing us to apply established insight from the
    adversarial community to bottom-up CG modeling, giving rise to new
    strategies for CG parameterization we term Adversarial-Residual-Coarse-Graining (ARCG).
    By doing so we facilitate the use of additional classes of CG model quality
    measures that may show promise in modifying the approximations
    characterizing the optimal CG model when using a constrained set of
    candidate potentials to represent the CG force-field.  
    We additionally find that it is possible to
    decouple the resolution at which one critiques the behavior of the CG model
    and the resolution at which a CG force-field is
    required: as an example we 
    describe a novel rigorous avenue to increase the expressiveness of bottom-up
    CG models through the use of virtual sites.  Critically, we do not utilize a
    full GAN architecture to generate CG samples; rather, we utilize the
    supporting theory
    \cite{goodfellow2014generative,reid2011,nowozin2016f,nguyen2010estimating,
    bouchacourt2016disco,bottou2017geometrical,arjovsky2017wasserstein} to
    optimize traditional CG force-fields.  

    In this work we discuss formal connections between CG and GAN-type implicit
    generative models and provide an initial implementation of the resulting ARCG 
    framework. Section \ref{theory_p} provides both an informal and a formal
    summary of the theoretical underpinnings, while section \ref{impl_p} provides
    details on a particular instance of ARCG and a public computational implementation.
    Section \ref{results_p} then
    provides results on three simple test systems, and section \ref{disc_p}
    outlines the consequences of the results and possible
    future studies. Section \ref{conc} provides concluding
    remarks.

\section{Theory} \label{theory_p}

    The purpose of this section is to both informally describe and formally
    define ARCG, and to summarize connections between ARCG and previous CG
    parameterization methods.  We begin by presenting an intuitive understanding
    of a specific form of ARCG to provide clarity for the subsequent mathematical
    description. We then follow by defining notation and the
    fine-grained/CG systems to which ARCG applies.  We define ARCG and
    describe its estimation and optimization. We then move to decouple the
    resolution at which one critiques the CG model from the resolution native to
    the CG Hamiltonian, thereby generalizing our application to systems
    containing virtual CG sites. We continue by discussing the corresponding
    challenges with momentum consistency, and we finish by summarizing ARCG's
    relationship to previous CG methods. 

\subsection{Informal Description of ARCG}

    Bottom-up CG models are parameterized to approximate the free energy surface
    implied by mapping fine-grained (FG) configurations to the CG
    resolution.\cite{noid2013perspective,saunders2013coarse}
    Generally, this entails considering many different possible CG models (each,
    for example, characterized by a different pair potential) and their
    relationship to the reference FG data. Often, this is
    operationalized by creating a variational statement and searching for the CG
    model that minimizes it (for example, minimizing the relative
    entropy between the CG model and FG data\cite{shell2008relative}). 
    After such a procedure is
    complete the modeler is well advised to visually inspect and compare the
    configurations produced by the selected CG model to
    those produced by the reference FG model. If the
    configurations are dissimilar, then the CG model is likely not adequate, and
    aspects of the variational statement or set of initial models considered
    must be modified and the parameterization process repeated. 

    It is natural to ask whether the final inspection of configurations produced
    by the FG and CG models can be intrinsically linked to the variational
    statement parameterizing the CG model. It is intuitive that for systematic CG
    parameterization methods derived from configurational consistency
    \cite{stillinger1970effective,lyubartsev1995calculation,izvekov2005multiscale,
    noid2007multiscale,noid2008multiscale,noid2008multiscale2,shell2008relative,
    mullinax2009generalized,karimi2011ibisco,
    carmichael2012new,dama2013theory,rudzinski2015bottom,lyubartsev2015systematic,
    vlcek2015rigorous}
    that when an indefinite amount of
    samples are used and all possible CG models are considered that the
    optimized CG model will perfectly reproduce the mapped FG
    statistics, and as a result, the configurations produced by
    the FG and CG models will be indistinguishable. 
    However, in cases where perfectly reproducing
    the FG statistics is infeasible it seems natural to ask if a model could
    be trained using this criteria of distinguishability directly.  

    While it could be possible in simple situations to use a human observer to
    intuitively rank CG models by considering the configurations they produce,
    this procedure quickly becomes subjective and untenable for complex models.
    A natural progression in method design is then to train a computer to
    distinguish CG models by comparing their samples against the reference data
    set. One appropriate statistical
    procedure is classification,\cite{james2013introduction} where a computer
    attempts to differentiate individual configurations based on whether they
    are more likely drawn from either the CG or mapped FG data sets. The implied
    procedure for CG parameterization is then to optimize the CG model such that
    it is intrinsically difficult to complete this task: as a result, the
    computer will inevitably make many mistakes on average when attempting to
    isolate configurations characteristic to only the FG and CG data. One
    possible intuitive manifestation of ARCG theory concretely implements
    this classification 
    procedure while maintaining clear connection to CG methods such as relative
    entropy minimization (REM).\cite{shell2008relative}  Previous CG
    parameterization methods have used similar, but not identical, motivations
    to produce parameterization
    strategies.\cite{lemke2017neural,vlcek2015rigorous,shell2008relative} ARCG theory serves to
    connect, clarify, and reframe these methods where possible while extending
    beyond the classification metaphor.

    It is important to note that the task of classification is a variational
    procedure itself:\cite{james2013introduction,reid2011} the ideal estimate of
    the true sources of a set of molecular configurations has a lower error than
    all other estimates. The optimization in classification searches over these
    various possible hypotheses. As a result, at each step of force-field
    optimization ARCG must perform this variational search over possible
    hypotheses, resulting in two nested variational statements in the full model
    optimization procedure: one required for classification, and the other
    for choosing the resulting CG model.  
    Importantly, the error rate of the optimal classifier
    can be explicitly linked to various $f$-divergences (e.g., relative entropy)
    evaluated between the mapped FG and CG distributions.\cite{reid2011}  This suggests an
    equivalent formalism with which to view ARCG: the variational estimation of
    divergences. This alternate interpretation additionally illustrates how
    additional divergences, such as the Wasserstein
    distance,\cite{arjovsky2017wasserstein} can be estimated,
    even without a clear connection to classification.  As a result, ARCG theory
    is primarily treated through the lens of variational divergence estimation 
    in the following sections.

    The variational estimation intrinsic to ARCG affords an interesting
    extension to traditional parameterization strategies: the resolution at
    which the CG Hamiltonian acts may be finer than the resolution at which the
    model is compared to the reference data. Equivalently, CG samples can be
    mapped before being compared to the mapped reference FG samples. For
    example, additional particles may be introduced to facilitate complex
    effective interactions between the CG particles, and then may be mapped out
    before comparing to the mapped reference FG samples.  Applying such a
    mapping creates issues with many other parameterization strategies as
    discussed in section \ref{related_methods}.

\subsection{Model Definitions and Selection}

    We consider a FG probability density $\refDensityMicro$ and a
    mapping operator $\map$ that maps a FG configuration to a
    CG configuration.  The FG simulation is constructed such that it
    produces samples from the Boltzmann distribution with respect to a
    FG Hamiltonian giving the following probability density:
    \begin{widetext}
        \begin{equation}
            \refDensityMicro(\fgpos^{3\fgnp},\fgmom^{3\fgnp}) 
            :=
            \refPFMicroPos^{-1}  \refPFMicroMom^{-1}
            \exp\left[
                -\beta \left(\sum_{i=1}^\fgnp
                \frac{\fgmom_i^2 }{2m_i} 
                + \refPotMicro(\fgpos^{3\fgnp})\right)
            \right]
             =
            \refDensityMicroMom(\fgmom^{3\fgnp})
            \refDensityMicroPos(\fgpos^{3\fgnp})
        \end{equation}
    \end{widetext}
    where $\beta$ is $\frac{1}{\textrm{k}_b T}$ with the temperature $T$
    set by the simulation protocol,
    $m_i$ are the FG masses, $\fgpos^{3\fgnp}$ and $\fgmom^{3\fgnp}$ are the
    FG positions and momenta variables, and
    our partition functions are defined as expected\footnote{
        Throughout this paper we omit proportionality constants related to
        indistinguishability and unit systems (including the factors of
        Planck's constant often introduced through quantum mechanical limits). The
        expressions used here can be considered in the context of dimensionless
        coordinates and distinguishable particles; reintroduction of these constants 
        is straightforward.}
        such that
    \begin{eqnarray}
        \refPFMicroPos
        &=
        &\int_{\refFeatDomPos}
        \exp\left[
            -\beta \refPotMicro(\fgpos^{3\fgnp})
        \right]
        \md \fgpos^{3\fgnp}
        \\*
        \refPFMicroMom
        &=
        &\int_{\refFeatDomMom}
        \exp\left[
            -\beta
            \sum_{i=1}^\fgnp
            \frac{\fgmom_i^2}{2m_i} 
        \right]
        \md \fgmom^{3\fgnp}
    \end{eqnarray}
    where the integrals are taken over the full domains of the position and
    momentum variables (denoted via $\refFeatDomPos$ and
    $\refFeatDomMom$).  The application of the CG map $\map$ produces CG
    configurations that follow an implied probability distribution.  $\map$ is
    constrained such that it is linear and can be decomposed into momentum and
    position components, i.e.,  \mbox{$\map(\fgpos^{3\fgnp},\fgmom^{3\fgnp}) = [
        \rmap(\fgpos^{3\fgnp});\pmap(\fgmom^{3\fgnp})
    ]$},\footnote{$\map$ is also
    typically\cite{noid2008multiscale} additionally constrained such that the
    resulting coordinates are linearly independent and
    unambiguously associate at most one atom to each CG
    site. These constraints are mostly unimportant to the
    work at hand except for when momentum consistency is considered, 
    but some care must be taken so that the
    corresponding densities exist.} implying a factorizable
    probability density $\refDensity( \pos^{3\cgnp} , \mom^{3\cgnp}) :=
    \refDensityPos( \pos^{3\cgnp} )\refDensityMom( \mom^{3\cgnp})$ over the CG
    variables defined as
    \begin{eqnarray}
        \label{mappingEqPos}
        \refDensityPos( \pos^{3\cgnp} )
        & :=
        &\int_{\refFeatDomPos} \refDensityMicroPos( \fgpos^{3\fgnp} )
        \delta (\rmap( \fgpos^{3\fgnp}) -  \pos^{3\cgnp} ) \md  \fgpos^{3\fgnp}
        \\*
        \label{mappingEqMom}
        \refDensityMom( \mom^{3\cgnp} )
        & :=
        &\int_{\refFeatDomMom} \refDensityMicroMom( \fgmom^{3\fgnp} )
        \delta (\pmap( \fgmom^{3\fgnp}) -  \mom^{3\cgnp} ) \md  \fgmom^{3\fgnp}.
    \end{eqnarray}
    Bottom-up CG models aim to directly produce samples from the
    distribution described by
    $\refDensity$.\cite{noid2007multiscale,noid2008multiscale}
    Ideally, this is achieved by defining a model CG
    Hamiltonian 
    $\left( \sum_{i=1}^\cgnp \frac{\mom_i^2}{2M_i} + 
    \cgPot(\pos^{3\cgnp})\right)$
    such that the corresponding Boltzmann statistics
    \begin{widetext}
        \begin{equation}
            \cgDensity(\pos^{3\cgnp},\mom^{3\cgnp})
            :=
            \cgPFPos^{-1}  \cgPFMom^{-1}
            \exp\left[
                -\beta \left(
                \sum_{i=1}^\cgnp
                \frac{\mom_i^2}{2M_i}
                +
                \cgPot(\pos^{3\cgnp}) \right)
            \right]
            =
            \cgDensityPos(\pos^{3\cgnp})
            \cgDensityMom(\mom^{3\cgnp})
        \end{equation}
    \end{widetext}
    are ideally identical to the mapped FG
    statistics, criteria expressed with the following CG consistency equations
    \cite{noid2008multiscale}
    \begin{eqnarray}
        \refDensityPos( \pos^{3\cgnp} )
        &= 
        &\cgDensityPos( \pos^{3\cgnp} )
        \label{fe_r}
        \\
        \refDensityMom( \mom^{3\cgnp} )
        &= 
        &\cgDensityMom( \mom^{3\cgnp} ).
        \label{fe_p}
    \end{eqnarray}
    Momentum and configurational consistency are generally treated separately, with
    momentum consistency exactly satisfied through direct definition of CG
    masses $M_i$ and configurational consistency approximated through a variational
    statement (as the corresponding integral is not
    generally tractable).\cite{noid2008multiscale} 
    We defer further discussion of momentum consistency until subsection
    \ref{momconst}.  The configurational variational statement is specific to
    the particular bottom-up CG method chosen and utilizes a variety of information
    depending on the method considered. Generally, knowledge of $\refPotMicro$,
    $\refPot$, and
    $\map$ are used. In many cases the 
    corresponding variational principle can be considered in the following 
    form
    \begin{equation} \label{qualityStatement}
        \cgParam^{\dagger} := 
        \argmin_{\cgParam} \Qual [ \cgDensityPosParam, \refDensityPos]
    \end{equation}
    where $\cgParam$ denotes the finite parameterization of our CG potential,
    $\cgParam^{\dagger}$ parameterizes our ideal model, and $\Qual$ is a function
    characterizing the quality of our model.
    Often,\cite{izvekov2005multiscale,shell2008relative,noid2007multiscale,
    noid2008multiscale,noid2008multiscale2,vlcek2015rigorous,stillinger1970effective,
    mullinax2009generalized} the exact form of the
    variational statement contains intractable integrals which are approximated
    via empirical averages from atomistic and coarse-grained trajectories.

    Importantly, while the models discussed in the remainder of this article fit
    into this framework, they differ in one important respect to many previous
    CG parameterization strategies: they introduce a variational
    definition of $\Qual$ itself, resulting in two nested variational statements
    in the numerical optimization procedure.

\subsection{Adversarial-Residual-Coarse-Grained Models}\label{icg}

    ARCG models are characterized by a set of possible $\Qual$ that are defined
    variationally as the difference in ensemble averages of a pair of coupled
    scalar functions. The functions, $f$
    \footnote{These functions, as well as other
    functions throughout the paper, must be integrable with respect to the
    measures defining the model or reference distributions. We will informally
    refer to integrable functions as functions. We do not, however, assume such
    functions are differentiable unless noted. Additionally, we will refer to
    functions which differ by measure zero as the same function when considering
    those which maximize expectation values.} 
    and $g$, are
    found as producing the maximum of the following variational definition
    \begin{equation} \label{icgresidual}
        \Qual[\cgDensityPosParam, \refDensityPos ] :=
        \max_{(f,g) \in \fpairspace}
        \left\{
            \langle f \rangle_\cgDensityParam - \langle g \rangle_\refDensity
        \right\},
    \end{equation}
    leading to a minimax variational statement for the fit model itself
    \begin{equation} \label{minMaxStatement}
        \cgParam^\dagger = \argmin_\cgParam
        \left[
        \max_{(f,g) \in \fpairspace}
        \left\{
            \langle f \rangle_{\cgDensityParam} - \langle g \rangle_\refDensity
        \right\}
        \right].
    \end{equation}
    In other words, for a specific choice of $\cgDensity$ and $\refDensity$ the
    numerical value of our residual is determined by a specific $(f,g)$ pair;
    all other choices of pairs of observables in $\fpairspace$ produce a more 
    optimistic estimate of the
    quality of our model. These observables are evaluated via their
    configurational average at the CG resolution. As we update $\cgParam$, the
    optimal choice of $(f,g)$ will change.  

    The definition of $\fpairspace$ depends on the particular $\Qual$ being
    specified. For example, if $f = g$ for all pairs in $\fpairspace$, this
    expression defines the class of Maximum Mean Discrepancy (MMD)
    distances,\cite{gretton2012kernel,dziugaite2015training} with each MMD distance then
    being defined by further constraints on $\fpairspace$. Typically, the 
    function space in MMD is restricted to the unit ball in a reproducing kernel
    Hilbert space, a choice which allows the maximization to be resolved 
    via a closed expression. The examples in this paper will estimate
    $f$-divergences: in this case, 
    \begin{equation}\label{pairspacelink}
        \fpairspace := 
        \left\{\left(-\oh l_\text{mod}\circ\hat{\eta}, \oh
        l_\text{mod}\circ\hat{\eta}\right) : 
        \hat{\eta} \in [0,1]^{\dom_\pos}
        \right\}
    \end{equation}
    where we have used $\circ$ to denote function compositions, e.g., \mbox{$f \circ g(x) :=
    f(g(x))$}, $[0,1]^{\dom_\pos}$ denotes the set of functions from $\dom_\pos$ to
    $[0,1]$, and $l_\text{ref}$ and $l_\text{mod}$ are functions determined by
    the specific $f$-divergence estimated and whose closed form is given in the next
    section.

    Model selection requires an optimization over $\cgParam$ to
    satisfy the external minimization in Eq. \eqref{minMaxStatement}. The
    strategies available for doing so depend on the structure of $\fpairspace$.
    For low dimensional parameterizations, it may be feasible to do a grid search
    over possible models and to use Eq. \eqref{icgresidual} to select the ideal
    model.  However, for higher dimensional parameter spaces an attractive
    option is to use methods utilizing the gradient with respect to $\cgParam$.
    If the maximized estimate over $\fpairspace$ is differentiable at a
    particular point with respect to $\cgParam$, then (due to the envelope
    theorem\cite{milgrom2002envelope}, see Appendix \ref{envthm}) the
    derivatives with respect to $\cgParam$ at that point only include terms
    related to the ensemble average over the model distribution, $\cgDensity$
    \begin{eqnarray} 
        \frac{\md}{\md \cgParam_i}
        \Qual[\cgDensityParam, \refDensity ] 
        &=
        &\frac{\md}{\md \cgParam_i}
        \max_{(f,g) \in \fpairspace}
        \left\{
            \langle f \rangle_{\cgDensityParam}
            - \langle g \rangle_\refDensity
        \right\}\;\;\;
        \\* 
        &=
        &\frac{\partial}{\partial \cgParam_i}
        \langle f^\dagger \rangle_{\cgDensityParam}
    \end{eqnarray}
    where $f^\dagger$ represents one of the optimal observables found at the
    internal maximum.  When the maximized inner estimate is expressible in
    closed form (which is true in the case of the $f$-divergences estimated in
    this paper), we can directly confirm the existence of this derivative.
    Assuming that the observable is regular enough such that the
    integral and derivative operators may be exchanged, simple
    substitution provides a covariance expression for estimation:
    \begin{equation}\label{covest}
        \frac{\partial}{\partial \cgParam_i}
        \left\langle f^\dagger \right\rangle_{\cgDensityParam}
        =
        \beta
        \left\langle f^\dagger \right\rangle_{\cgDensityParam}
        \left\langle
        \frac{
            \partial U_\cgParam
        }
        {
            \partial \cgParam_i
        }
        \right\rangle_{\cgDensityParam}
        - \beta
        \left\langle
        f^\dagger
        \frac{
            \partial U_\cgParam
        }
        {
            \partial \cgParam_i
        }
        \right\rangle_{\cgDensityParam}.
    \end{equation}
    These results suggest a straightforward numerical optimization of Eq.
    \eqref{minMaxStatement} using gradient descent and related first order
    methods (e.g., RMSprop\cite{hinton2012neural}).  We represent $\fpairspace$
    by indexing with a finite dimensional vector $\varParam$. At each iteration
    of optimization, holding $\cgParam$ constant, we maximize over $\varParam$
    using samples from the model and reference distributions to estimate our
    expected values; then, holding $\varParam$ constant, we take a single step
    on the gradient of $\cgParam$ estimated by the sample average of the
    covariance expression. This two step process is completed until convergence
    of $\cgParam$. 

    Not all definitions of $\fpairspace$ produce meaningful procedures for
    creating CG models.  Generally, particular forms of $\Qual$ are
    derived individually, each of which is amenable to the procedures outlined
    here.  We continue by investigating an informative subset of possible
    $\Qual$, characterized via $f$-divergences, that will provide functionality
    directly encompassing REM CG,\cite{shell2008relative} as well as 
    previous approaches by \citet{stillinger1970effective} 
    and \citet{vlcek2015rigorous}.

\subsection{$f$-divergences}\label{fdiv}

    The $f$-divergences are a category of functions characterizing the
    difference between two distributions.\cite{reid2011}
    When probability density functions are
    available we can express this family of divergences as
    \begin{equation}
        \fd_f(\refDensity,\cgDensity) := 
        \int_\chi \cgDensity(x) 
        f \left( \frac{\refDensity(x)}{\cgDensity(x)} \right) \md x
    \end{equation}
    where each member of the family is indexed by a convex
    function $f$ that
    satisfies $f(1) = 0$. Relative
    entropy, the divergence central to REM CG, 
    can be obtained by defining
    \mbox{$f(x) := x \log x$},\footnote{We note that the $x$ proceeding the $\log$ here
    effectively changes the distribution over which $\log$ is averaged; relative
    entropy traditionally averages $\log \left( \frac{\refDensity(x)}{\cgDensity(x)}
    \right)$ over $\refDensity$.} 
    and the Hellinger distance, central to previous methods
    by \citet{stillinger1970effective} and \citet{vlcek2015rigorous}
    can be obtained by via \mbox{$f(x) := \left(\sqrt{x}-1\right)^2$}.

    The $f$-divergence between $\cgDensity$ and $\refDensity$ can be
    expressed in multiple variational statements.\cite{reid2011,
    nowozin2016f,nguyen2010estimating,bottou2017geometrical} We here utilize its
    duality with the difficulty of classification which is
    mathematically expressed in the following formulation, giving the 
    form
    \begin{widetext}
        \begin{equation} \label{decision_fdiv}
            \fd_f(\cgDensity,\refDensity) =
            \max_{\hat{\eta}\in [0,1]^{\dom_\pos}}
            \left[
            -
                \oh
                \langle
                    l_\text{mod} \circ \hat{\eta}
                \rangle_\cgDensity
            -
                \oh
                \langle
                    l_\text{ref} \circ \hat{\eta}
                \rangle_\refDensity
            \right]
        \end{equation}
    \end{widetext}
    where
    \begin{eqnarray*}\label{decision_fdiv_losses}
        \bloss(x) 
        & :=
        & - 2(1-x)
        f \left(
            \frac{x}{1-x}
        \right)
        \\
        l_\text{mod}(h) 
        & := 
        &\bloss(h) - h \left. \frac{\partial\bloss}{\partial x}\right\rvert_h
        \\
        l_\text{ref}(h) 
        & := 
        & \bloss(h) + 
        (1-h) \left. \frac{\partial\bloss}{\partial x}\right\rvert_h.
    \end{eqnarray*}
    The function $\hat{\eta}$ is a function of a CG configuration,
    mapping each configuration to a real number in $[0,1]$.\footnote{We note
    that certain proper losses may only be defined on $(0,1]$, $[0,1)$, or
    $[0,1)$. The is partially reflective of the fact that certain
    $f$-divergences, such as relative entropy, are not always defined when the
    support of the corresponding densities is not the same. There exist cases
    such that the limiting behavior of such losses is still valid for
    distributions with differing support, such as the log loss. The optimization
    performed in this paper only compares distributions for which the KL
    divergence is defined, and for which $\eta \in (0,1)$.
    The variational statements hold more generally; see
    \citet{reid2011} for more details.}  Note that substitution into Eq.
    \eqref{minMaxStatement} (along with the removal of prefactors) provides us
    with our training residual
    \begin{equation} \label{hedgebet_minMaxStatement}
        \cgParam^\dagger =
        \argmin_\cgParam
        \left[
            \max_{\hat{\eta}\in [0,1]^{\dom_\pos}}
        \left\{
            -
            \langle
                l_\text{mod} \circ \hat{\eta}
            \rangle_{\cgDensityParam}
            -
            \langle
                l_\text{ref} \circ \hat{\eta}
            \rangle_\refDensity
        \right\}
        \right]
    \end{equation}
    where the optimal $\hat{\eta}$ producing the
    corresponding $f$-divergence, denoted $\eta$, is known to
    be\cite{reid2011}
    \begin{equation}\label{etadef}
        \eta(x) = \frac{\refDensity(x)}{\cgDensity(x) + \refDensity(x)}.
    \end{equation}
    While here we have denoted our inner variational statement as optimizing
    over a space of functions instead of pairs of functions, this is equivalent to Eq. \eqref{minMaxStatement}
    when defining $\fpairspace$ via Eq. $\eqref{pairspacelink}$.
    In the context of $f$-divergences, when $\fpairspace$ contains $(-
    \oh l_\text{mod}\circ\eta, \oh l_\text{mod}\circ\eta)$, i.e. when optimization with
    the population averages would return the corresponding $f$-divergence, we
    will refer to that $\fpairspace$ as being \emph{complete}.  Additionally,
    when $\fpairspace$ is expressive enough such that it is \emph{complete} for
    each step in an optimization process, we will additionally refer to it as
    \emph{complete}, with the distinction evident from context. We provide
    concrete expressions for calculating relative entropy in section
    \ref{impl_p} and in appendix \ref{loss_derivation}.

    Despite the seemingly opaque form of Eq. \eqref{hedgebet_minMaxStatement},
    the variational statement provided has a notable intuitive description,
    which will be useful when considering implementation and connections to
    similar methods.  Consider an external observer that has access to a mixture
    of molecular configurational samples, some of which are produced by our
    mapped reference simulation and others from our CG model (termed our
    reference and model samples, respectively). The observer is faced with the
    following task: they must distinguish which examples came from which source
    based solely on configurational details. We represent the observer's guess by the
    function $\hat{\eta}$, which maps each molecular configuration to a
    number in the interval $[0,1]$. We associate the label 0 with configurations
    from our model and the label 1 with configurations from our reference set
    (note that the labels are discrete, but our estimate is a number between 0
    and 1 inclusive). We decide in this case to use the square loss, giving us the
    following definitions for our loss functions:
    \begin{equation}
        l^{\text{sq}}_\text{mod}\circ\hat{\eta}(x) := \hat{\eta}(x)^2
    \end{equation}
    \begin{equation}
        l^{\text{sq}}_\text{ref}\circ\hat{\eta}(x) := (1-\hat{\eta}(x))^2
    \end{equation}
    where $x$ is a particular molecular configuration.
    For example, if the observer guesses a probability of 0.68 for a
    configuration that was drawn from the reference set, they are penalized
    $(1-0.68)^2=0.1024$. If the configuration instead came from the model data
    set, they are penalized $0.68^2=0.4624$. The observer wishes to minimize
    their penalty, and if they are able to guess 1 for all configurations drawn
    from the reference set and 0 for all the configurations drawn from the model
    set, then their loss will be minimized at 0.

    If the model is very poor, achieving a average loss of 0 will be easy---the
    configurations from the model will be distinct from the reference
    configurations.  However, for higher quality models many of their
    configurations will plausibly come from either the model or the reference
    simulation.  
    Even with the perfect
    $\hat{\eta}$, a configuration which has a
    $50\%$ probability of coming from the reference and model sets will entail a
    minimum loss of 0.25 (this minimum is entailed when the estimated
    probability is \emph{also} 50\%); this loss cannot be reduced further. We refer to this loss as
    the irreducible loss. This is analogous to the least squares residual present in linear
    regression with Gaussian noise.  The ideal line minimizes the
    least squares residual, but the least squares residual is nonzero as the
    line cannot perfectly fit the data.

    This inability to perfectly distinguish samples is directly related to our
    $f$-divergences (e.g., relative entropy).\cite{reid2011} Modifying the
    manner in which we penalize incorrect predictions (via $l_\text{mod}$ and
    $l_\text{ref}$) specifies which divergence is produced. In this example, we
    have decided on the form of our losses directly; when estimating a
    particular $f$-divergence the expressions defining the losses are given by
    Eq. \eqref{decision_fdiv}.  This loss function is asymmetric depending on
    the true origin of the sample: $l_\text{mod}$ penalizes a prediction on a
    sample gained from the model, while $l_\text{ref}$ penalizes a prediction on
    a reference sample.  Notably, while there are constraints on what
    functions $l_\text{mod}$ and $l_\text{ref}$ can be defined as in order for
    $\eta$ (the optimal $\hat{\eta}$) to obey Eq. \eqref{etadef}, these
    constraints are already taken into account by Eq. \eqref{decision_fdiv}: a
    valid $f$-divergence will always yield losses whose optimal estimate is
    given by Eq.  \eqref{etadef}.

    As a result, we simply need to train a classifier with a loss on our samples
    and consider the average loss implied by its probabilistic predictions.  An
    extended formal description of this task and the corresponding duality is 
    presented in \citet{reid2011}. This interpretation is
    central to the term adversary in the name of Generative Adversarial
    Networks:\cite{goodfellow2014generative} the adversary attempts identify the
    source of each sample and we wish to make its task as difficult as possible.

\subsection{Virtual Sites}\label{vs}

    The ARCG framework can be lightly generalized to decouple the resolution at
    which the CG potential acts and the resolution at which we compare our CG
    and reference systems. More specifically, we see that we can apply a
    distinct mapping operator to our CG system before it is compared to the
    mapped FG samples. To better illustrate the practical use of this
    extension we begin by providing a motivating example.

\begin{figure}[h]
        \centering
        \includegraphics[width=8cm]{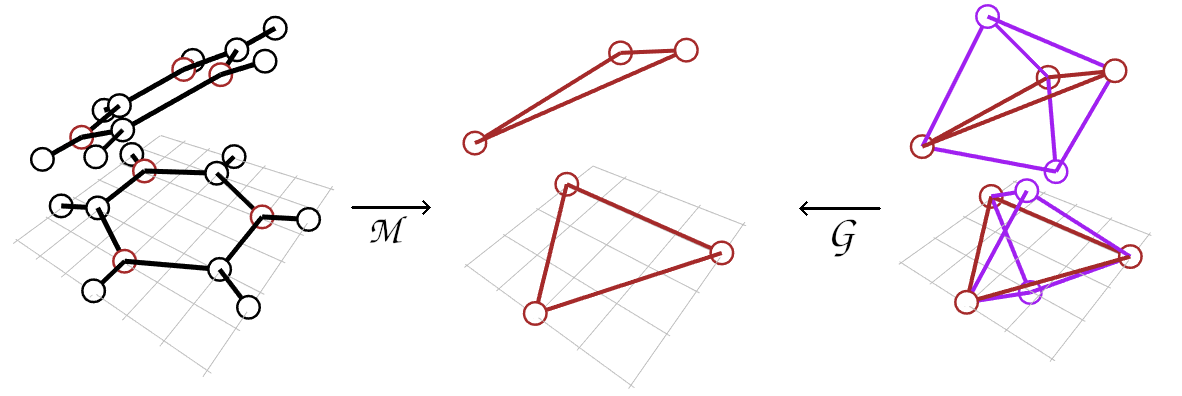}
        \caption{An example of virtual particle usage. The
        atomistic representation of benzene (left) is
        mapped via $\map$ to a CG representation (center) only preserving three
        carbons (red). The full CG representation
        (right) of the same
        configuration has these three carbons and two
        additional virtual sites (purple) to help a pairwise
        potential capture the correct PMF. These sites are removed
        upon application of the virtual particle map $\vmap$. These virtual
        sites have no atomistic counterpart.}
        \label{fig:vsite}
    \end{figure}

    As previously discussed, many bottom-up CG methods are shown to produce the
    ideal PMF when they are allowed to adopt any force-field in the ideal
    sampling limit. However, CG models are often limited to molecular mechanics
    type potentials (e.g., pairwise
    nonbonded potentials), which often do not contain the ideal PMF as a possible
    parameterization. For example, one might use Multiscale
    Coarse-Graining\cite{izvekov2005multiscale,noid2007multiscale,noid2008multiscale,
    noid2008multiscale2} (MS-CG) to parameterize a CG
    lipid bilayer in which all of the solvent and some of the lipid degrees of
    freedom have been removed. Upon generating samples using the CG model we may
    find that certain properties of the membrane, such as its thermodynamic
    force of bilayer assembly, are poor.  However, the MS-CG method has likely
    provided one with its correct characteristic approximation; in order to
    improve the model with the same parameterization method one must either
    increase the complexity of the CG force-field via higher
    order terms or retain more FG details
    via modification of
    $\map$, the CG map seen in Eqs. \eqref{mappingEqPos} and
    \eqref{mappingEqMom}.  Here, we discuss a third option: 
    augmenting the CG representation
    directly without modifying $\map$.  
    As a simple example consider modeling the
    interaction of two benzene molecules using a CG pairwise
    potential. The CG representation is given by three sites
    per benzene ring. It may be difficult to capture the $\pi$-stacking effect
    using this type of potential at the CG resolution.
    As a remedy one could add particles
    normal to the plane containing the benzene molecule, as
    shown in fig. \ref{fig:vsite},
    without associating these additional CG sites to FG
    sites via $\map$. Importantly, however, we
    will only critique the behavior of our CG model after these virtual sites
    have been integrated out: the CG model is optimized to minimize the
    relative entropy between the mapped FG reference and CG model after the
    integration over the possible positions of these virtual CG sites.

\begin{figure}[h]
        \centering
        \includegraphics[width=8cm]{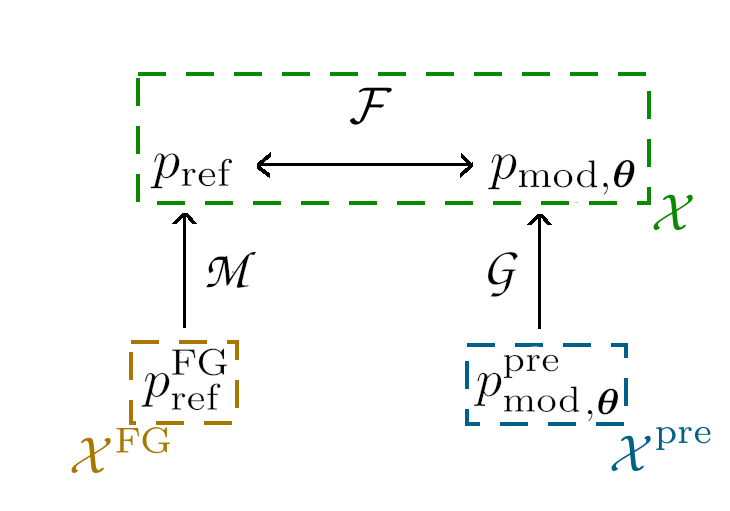}
        \caption{The relationship between resolutions when comparing FG and CG
        systems at a custom resolution, such as the case of virtual sites.
        Samples from the pre-CG domain $\preFeatDom$ (e.g., a CG configuration
        including virtual sites) are mapped to the CG domain $\dom$ (e.g., a CG
        configuration without virtual sites) via $\vmap$; samples from the FG
        domain $\refFeatDom$ (e.g., atomistic) are mapped to the same CG domain $\dom$
        via $\map$. The mapped samples are then compared via $\Qual$.}
        \label{fig:resdiag}
    \end{figure}

    Description of the formalism encompassing these situations requires us to
    suitably expand our notation. We still consider all distributions described
    previously but use the following modifications: first, samples from
    $\cgDensity$ are no longer generated by a simulation using the
    approximated PMF as its Hamiltonian. Instead, these samples are produced via
    a new mapping operator $\vmap$ and simulation of a new finer grained representation
    characterized by $\preDensity$ via its own Hamiltonian
    $\left( \sum_{i=1}^\prenp \premom_i^2 / 2\premass_i + \prePot(\prepos^{3\prenp})\right)$ 
    where $\premass_i$ are the masses at the pre-CG resolution.
    As a result, $\cgDensity$ is redefined with the following relations.
    \begin{eqnarray}
        \cgDensityPos( \pos^{3\cgnp} )
        &:=
        &\int_{\preFeatDomPos} \preDensityPos( \prepos^{3\prenp} )
        \delta (\rvmap( \prepos^{3\prenp}) -  \pos^{3\cgnp} ) \md  \prepos^{3\prenp}
        \;\;\;
        \label{fe_r_pre}
        \\*
        \cgDensityMom( \mom^{3\cgnp} )
        &:=
        &\int_{\preFeatDomMom} \preDensityMom( \premom^{3\prenp} )
        \delta (\pvmap( \premom^{3\prenp}) -  \mom^{3\cgnp} ) \md  \premom^{3\prenp}
        \;\;\;
        \label{fe_p_pre}
    \end{eqnarray}
    The resulting relations between resolutions are summarized in fig.
    \ref{fig:resdiag}.

    Importantly, our training procedure needs two minor modifications. First, the
    variational estimation of divergences presented in Eq. \eqref{icgresidual} is
    composed solely of ensemble averages, which are approximated via sample
    averages; these averages can be evaluated by generating empirical
    samples from $\cgDensity$ via samples drawn from
    $\preDensity$ 
    and application of $\vmap$. This is a consequence of 
    Eq. \eqref{ensemble_av_id}.
    \begin{equation}\label{ensemble_av_id}
        \langle
            f
        \rangle_\cgDensity
        =
        \langle
            f \circ \vmap
        \rangle_\preDensity
    \end{equation}
    Second, the gradients
    required for optimization of the parameters of the
    variational search ($\cgParam$) are calculable again through 
    Eq. \eqref{ensemble_av_id},
    allowing us to utilize our previous expression Eq. \eqref{covest} at the
    resolution native to our new pre-CG Hamiltonian by minimizing the
    variationally optimized observable composed with $\vmap$.

    Importantly, while our examples in this section have primarily concerned
    situations in which fictional particles are added to the CG representation
    and then completely integrated over before calculating divergences, $\vmap$ can
    easily be generalized. Fundamentally, it has the full flexibility of $\map$;
    similarly, additional constraints are born from maintaining momentum
    consistency via methods described in the next subsection. However, if one
    discards momentum consistency, it is possible to maintain an intuitive
    pre-CG representation while nonlinearly modifying $\map$ and $\vmap$ to
    represent custom high-dimensional observables. In this case these mapped
    distributions are used for determining the quality of the pre-CG
    model. We reserve the bulk of our discussion and investigation of this more
    complex option to a future article.

\subsubsection{Momentum Consistency}\label{momconst}

    Previous sections have discussed the configurational variational statement
    central to ARCG; here, we discuss how to ensure momentum consistency.
    In the case that no pre-CG resolution is considered, momentum
    consistency in ARCG may be achieved through identical methods as stated in
    previous approaches, such as MS-CG.\cite{noid2008multiscale}  However, when
    considering three distinct resolutions momentum consistency takes on a
    slightly modified form. We provide suitable constraints for a common case
    below, although extensions are straightforward. 

    Momentum consistency is characterized by the following equation:
    \begin{equation}\label{fe_p}
        \refDensityMom( \mom^{3\cgnp} )
        = \cgDensityMom( \mom^{3\cgnp} ).
    \end{equation}
    We here consider the specific case where both $\rmap$ and
    $\rvmap$ are linear functions that satisfy the constraints defined in the
    MS-CG work:\cite{noid2008multiscale} $\rvmap$ is limited to associate each CG
    site in $\preFeatDom$ unambiguously to at most a single site in $\dom$ and has
    imposed translational  and positivity constraints, and analogous 
    constraints are placed on
    $\rmap$ (see appendix for more details). 
    The momentum map $\pmap$ (and $\pvmap$ with appropriate
    modifications) is assumed to
    take the following form as in reference
    \citenum{noid2008multiscale}:
    \begin{equation}
        \pmap_I(\fgmom^{3\fgnp}) := M^\map_I
        \sum_{i\in \mathcal{I}^\map_I}
        \frac{{c^\map_{Ii}}^2 \fgmom_i}{m_i},
    \end{equation}
    In this case, previous work
    \cite{noid2008multiscale} has shown 
    that the constants defining $\pmap$ (and similarly
    $\pvmap$) can be combined
    with the masses of the sites contributing to a mapped
    site to provide a
    definition of the mapped masses (Eq. \eqref{simplemom})
    that define a Boltzmann 
    distribution equal to the mapped momentum distribution
    \begin{equation}\label{simplemom}
        \left({M^\map_I}\right)^{-1} := \sum_{i\in \mathcal{I}^\map_I}
        \frac{{c^\map_{Ii}}^2}{m_i},
    \end{equation}
    where ${M^\map_I}$ is the mass of CG particle $I$ as
    implied by map $\map$, $\mathcal{I}^\map_I$ is the set
    of all atoms that map to CG site $I$ according to map
    $\map$, and $c^\map_{Ii}$ is the coefficient describing
    how the positions of FG particle $i$ contribute to CG
    particle $I$ according to map $\map$.  More generally,
    this implies that we can explicitly characterize the
    mapped momentum distributions for both the mapped FG and
    mapped CG systems, which when combined with Eq. \eqref{fe_p}
    provides the following relation implying momentum
    consistency in a system with virtual particles
    \begin{eqnarray}
        \exp\left(- \beta \sum_{I=1}^N \frac{\mom_I^2}{2 M^\vmap_I} \right)
        &\propto&
        \exp\left(- \beta \sum_{I=1}^N \frac{\mom_I^2}{2 M^\map_I} \right)
        \\
        \left({M^\vmap_I}\right)^{-1} &:=& \sum_{i\in \mathcal{I}^\vmap_I}
        \frac{{c^\vmap_{Ii}}^2}{\premass_i}.
    \end{eqnarray}
    The only solution to this equation is to set $M^\vmap_I = M^\map_I$ for each CG site
    $I$; in this case we find a set of equations implying
    consistency (Eq. \eqref{mt_mass_crit}).
    \begin{equation}\label{mt_mass_crit}
        \left[
            0
            =
             \sum_{i\in \mathcal{I}^\map_I}
            \frac{{c^\map_{Ii}}^2}{m_i}
            -
            \sum_{i\in \mathcal{I}^\vmap_I}
            \frac{{c^\vmap_{Ii}}^2}{\premass_i}
        \right]
        \forall \text{ CG sites }I
    \end{equation}
    Note that these equations are positively constrained with respect to masses
    and mapping constants (along with the previously stated constraints).
    This provides a simple condition connecting our FG masses,
    pre-CG masses, $\map$, and $\vmap$, and allows one to check for momentum
    consistency if all the relevant variables are defined.  
    It is important to note that $I$ indexes the
    CG sites at the resolution of $\refDensity$ and
    $\cgDensity$---that is, without the virtual particles. As such, in the
    case of $\vmap$ simply dropping virtual particles consistency is
    trivially satisfied by simply matching the masses of the non dropped
    particles to those implied by the FG system with $\map$. Additional details
    may be found in the appendix.

\subsection{Related Methods}\label{related_methods}

    Despite differences in representation, ARCG can be formulated to elucidate
    connections to a variety of previous CG parameterization strategies, some of
    which have been mentioned in previous sections.  This is performed via the
    appropriate design of the characteristic function space
    $\fpairspace$ in Eq. \eqref{icgresidual}.
    Additionally, ARCG bears resemblance to a recent CG method based
    on distinguishability and classification.\cite{lemke2017neural} In this
    section we make explicit connections between the $f$-divergence implementation
    presented in this article and such external methods.  
    The applications of the $f$-divergence duality presented here are
    in the infinite sampling limit with a fully expressive variational search;
    in practice, significant differences in seemingly equivalent methods may
    arise.

    Classification has been recently used to train a CG model by using the
    resulting decision function $\hat{\eta}^\dagger$ to directly update the CG
    configurational free energy.\cite{lemke2017neural}  This is motivated by
    noticing that the $\eta$ that satisfies the variational bound in
    Eq. \eqref{decision_fdiv} can be related to the
    pointwise free energy difference as
    \begin{equation}
        \log \frac{1-\eta}{\eta} = \log \refDensity - \log \cgDensity,
    \end{equation}
    suggesting a procedure where 
    $\log (1-\hat{\eta})- \log(\hat{\eta})$ is
    scaled and used as an additive update to the CG potential.  This procedure
    is similarly valid using any of the $f$-divergence losses discussed in this 
    article.\cite{reid2011}  However, beyond the
    differing update rules, the variational divergence approach presented in
    this article is differentiated by a subtle but important difference in
    characteristic assumptions.  The divergence interpretations of ARCG 
    rely on the completeness of $\fpairspace$, but place no constraint on
    $\{\cgDensityParam\}_{\cgParam\in\mathbf{\Theta}}$, where $\mathbf{\Theta}$ 
    denotes the set of all model parameterizations considered.  
    In contrast, the interpretation of the method of
    \citet{lemke2017neural} also requires an fully expressive $\fpairspace$;
    however, as the update to $\cgDensity$ inherently utilizes members of
    $\fpairspace$, the method naturally also forces 
    $\{\cgDensityParam\}_{\cgParam\in\mathbf{\Theta}}$ to be fully expressive,
    i.e. $\refDensity \in \{\cgDensityParam\}_{\cgParam\in\mathbf{\Theta}}$. In other words,
    $\fpairspace$ and $\{\cgDensityParam\}_{\cgParam\in\mathbf{\Theta}}$ are directly coupled.
    As a result, in the case that the classifier used in the additive update
    method similarly has a relation to a specific
    $f$-divergence, an ideal model would always be chosen, rendering the specific
    choice of $f$-divergence inconsequential.  Beyond this it is unclear how to
    expand the update rule of \citet{lemke2017neural} to apply to virtual sites, as the
    classifier is only directly present at the resolution of $\cgDensity$ and
    extension of the update to the resolution of $\preDensity$ is unclear.

    REM CG proposes that approximate CG models should be parameterized by
    minimizing the relative entropy,\cite{shell2008relative} or KL-divergence, between the
    distributions produced at the FG resolution:
    \begin{equation} \label{recg_fg}
        \int_\refFeatDom \refDensityMicro (x)
        \log \left(
            \frac{\refDensityMicro(x)}
            {p^\text{FG}_\text{mod}(x)}
        \right)
        \md x
    \end{equation}
    where we have introduced  a new quantity, $p^\text{FG}_\text{mod}$,
    defined to be the probability density implied by the CG model over
    FG space (which is not used in ARCG theory); the exact form implied over the
    FG space depends on the interpretation of REM CG
    considered.\cite{rudzinski2011coarse}  
    This differs by a constant (when considering CG force-field
    optimization) from the relative entropy
    considered at resolution of the CG model, given by
    \begin{equation}\label{recg_cg}
        \int_\dom \refDensity(x)
        \log \left(
            \frac{\refDensity(x)}
            {\cgDensity(x)}
        \right)
        \md x.
    \end{equation}
    KL-divergence is an $f$-divergence (generated by $f(x): = x \log x$) 
    and in the case of Eq. \eqref{recg_cg}
    can resultingly be formulated and solved for in the current framework,
    providing the following losses through Eq. \eqref{decision_fdiv}
    \begin{eqnarray}\label{re_losses}
        l^{\text{RE}}_{\text{ref}}(h) 
        &=
        &2 \left[\log \left( \frac{1-h}{h} \right)-1\right]
        \\*
        l^{\text{RE}}_{\text{mod}}(h)
        &=
        & 2 \frac {h} {1-h}.
    \end{eqnarray}
    We utilize this method for the computational examples presented in Sec.
    \ref{impl_p}. We note that the full specification of REM CG considers
    comparing a coarser CG model to a finer FG model at the FG resolution by
    defining a new model density at the FG resolution
    ($p^\text{FG}_\text{mod}$), as where we have used many-to-one
    functions to reduce the resolution of the FG and pre-CG model in our
    theoretical approach.  However, calculation
    of the relative entropy at CG resolution produces the same model selection
    rule as the FG relative entropy when considering the CG force-field.
    Optimization of systems with virtual particles is not straightforward via
    REM CG as most refinement schemes require $\langle\partial_\cgParam
    \cgPot_\cgParam\rangle$ which is difficult to calculate as the explicit form
    of $\cgPot_\cgParam$ is unknown in the case of virtual particles.

    \citet{schoberl2017predictive} extended REM CG by framing coarse-graining as
    a generative process where the FG statistics are non-deterministically
    produced by the CG variables by means of a backmapping operator, a method
    termed Predictive Coarse-Graining (PCG).  This approach allows optimization
    of the backmapping operator itself and additionally allows more
    flexibility in describing the connection between the FG and CG systems.
    This allows PCG to describe CG models with
    virtual particles.  Additionally, PCG is trained using
    expectation-maximization, which can be framed as a two part process with a
    variational search providing the information for a gradient update of the
    parameters. PCG differs from ARCG in multiple ways. First, PCG aims to
    optimize an iteratively tightened lower bound on the relative entropy of the
    CG model, whereas ARCG encompasses the optimization of a larger variety of
    possible metrics, including relative entropy.  Additionally, the variational
    estimation in PCG is solved via a closed form expression and generates a
    gradient update which optimizes said lower bound, as where
    the variational optimization in ARCG is solved iteratively in practice and
    provides the exact gradient of relative entropy. Finally, ARCG is not
    formulated as generating statistics at the FG resolution and instead is
    formulated on the CG resolution. Despite these differences, the overall
    similarity between PCG and ARCG suggests that the two methods could be used
    to extend each other.  We reserve a detailed analysis of these connections
    for a future work.

    Alternatively, recent work by \citet{vlcek2015rigorous} (as well as previous work by
    \citet{stillinger1970effective}) suggests that the 
    Bhattacharyya distance ($BD$) Eq. \eqref{bhat_distance}
    is a natural metric to judge approximate models. 
    \begin{eqnarray}
        BC(\cgDensity,\refDensity)
        &:=
        &\int_\dom \sqrt{\cgDensity(x) \refDensity(x)} \md x
        \label{bhat_coef}
        \\*
        BD(\cgDensity,\refDensity) 
        &:= 
        &- \log BC(\cgDensity,\refDensity)
        \label{bhat_distance}
    \end{eqnarray}
    While the Bhattacharyya distance is not an
    $f$-divergence, it is related to one via a monotonic transformation: the
    Hellinger distance ($H$)
    \begin{eqnarray}\label{hellinger_d}
        H(\cgDensity,\refDensity) 
        &:= 
        &\sqrt{1-BC(\cgDensity,\refDensity)}
        \\*
        &= 
        &\fd_{\left(\sqrt{t}-1\right)^2}(\cgDensity,\refDensity).
    \end{eqnarray}
    This can be variationally approximated in the same framework
    as REM CG, resulting in the following losses:
    \begin{eqnarray}\label{hellinger_losses}
        l^H_{\text{mod}}(h) 
        &=
        &
        2 
        \sqrt{
            \frac
            {h}
            {1-h}
        }
        \\*
        l^H_{\text{ref}}(h)
        &=
        &2 
        \sqrt{
            \frac{1-h}{h}
        }.
    \end{eqnarray}
    Justification of the Bhattacharyya distance may be grounded in
    information geometry and the distinguishability of samples produced by the
    FG and CG models. Despite the apparent similarity to the fictional
    game described earlier, the justification of \citet{vlcek2015rigorous} is
    grounded in distinguishing populations via their collective empirical
    samples, while our game focuses on distinguishing individual configurations.
    The stated connection simply occurs through our duality with
    $f$-divergences.

    Inverse Monte Carlo (IMC),\cite{lyubartsev1995calculation} also known as
    Newton Inversion (NI), minimizes an observable that characterizes the
    difference between the mapped FG and CG systems (often through their radial
    distribution functions) and may be used on systems with virtual particles.
    The distributions utilized for this comparison are often low dimensional and
    are calculated via traditional binning approaches. ARCG may be viewed
    similarly as minimizing the expected value of observables; however in ARCG
    the observable minimized at each step of optimization must be variationally
    found, and subsequently changes from step to step. However, due to the
    envelope theorem, the derivatives calculated for both ARCG and IMC/NI share
    a similar covariance form shown in Eq. \eqref{covest}. Additionally, the
    typical approach in IMC/NI requires histograms to calculate the desired
    empirical correlation functions, limiting the metric to low dimensional
    distributions; ARCG does not perform binning of any kind.

    There exist additional CG methods that are difficult to directly compare to
    ARCG (e.g., references \citenum{izvekov2005multiscale,mullinax2009generalized,
    lyubartsev1995calculation,noid2008multiscale,noid2007multiscale,noid2008multiscale2}). 
    However, in general, most methods considered
    make assumptions that strongly inhibit virtual site application.
    Specifically, methods often assume that the CG potential (or its
    derivatives) can be applied at
    the resolution of the CG samples acquired (either through calculation of the
    residual or the update strategy facilitating optimization), although
    extensions are sometimes feasible. For example, traditional MS-CG
    force-matching optimizes the CG force-field to optimally match mapped
    forces; with a general linear $\vmap$ and $\prePot$ this would likely
    require an iterative procedure to determine the mean force implied at the CG
    resolution by $\vmap$ and $\prePot$.  Alternatively,
    gYBG inverts two- and
    three-body CG correlations to produce a force-field at the corresponding
    resolution of the observed correlations; similarly, Iterative Boltzmann
    Inversion requires a map to define the iterations that connect
    modifications in the potential to changes in the observed correlations
    (which is nonintuitive when considering parameters associated with general
    virtual sites). These limitations often do not appear to be fundamental ones,
    but rather one of implementation; extensions to these
    methods that
    circumvent this limitation are likely possible.  There are three
    straightforward strategies to remove this limitation,
    the first two of which the
    authors know are in current use. First, several methods such as binning or kernel
    density estimation are used to approximate the probability density at a
    resolution differing from the CG configurational Hamiltonian (e.g., the
    radial distribution approach in reference \citenum{vlcek2015rigorous}). This
    approach is often limited to lower dimensional spaces when comparing models.
    Second, constraints are placed on virtual sites such that $\prePot$ may be
    related via closed expression to $\cgPot$.\cite{jinvirtual}
     This approach inherently requires
    limiting the type of virtual site considered.  Third, methods
    that allow the observed mapped FG sample to be backmapped to the pre-CG
    domain are applied and then traditional approaches are used on the
    backmapped sample. In contrast, ARCG is well suited to higher dimensions,
    imposes no constraint on the virtual sites, and does not require
    backmapping; however, it incurs increased training complexity.

    Finally, we note that while there is significant overlap between ARCG and
    GANs with respect to the residual calculation and optimization, the method
    by which samples are produced in the models is conceptually distinct. GANs
    are characterized by transforming noise to a fit a desired 
    distribution; the optimization of the model parameters modifies the nature
    of this transformation. In contrast, the transformation present in ARCG is
    held constant, while the underlying sample generating process is modified.

\section{Implementation} \label{impl_p}

    Previous sections have provided abstract descriptions of the ARCG method,
    including the specific form with connection to $f$-divergences. In this
    section we provide the corresponding concrete expressions for optimizing
    models using relative entropy by implementing the classification based
    approach described in Sec. \ref{fdiv}.  Additional practical points on
    implementation, relaxations of the method for stability, and the
    specification of $\fpairspace$ are also discussed.

    As previously noted, the relative entropy between $\refDensity$ and
    $\cgDensity$ is an $f$-divergence and is obtained by setting $f(x) := x
    \log x$.  This implies equivalence with a classification task
    with the aforementioned losses in Eq. \eqref{re_losses}, from which we derive the
    model optimization statement using Eq. \eqref{hedgebet_minMaxStatement} and
    associated gradients using Eq. \eqref{covest}, such that
    \begin{widetext}
        \begin{equation}\label{relentlearn}
                \Qual^{\text{RE}}
                \left[
                    \preDensity_\cgParam,\refDensity; \vmap
                \right]
            =
            \max_{\hat{\eta}}
            \left\{
                 - \left\langle
                    \log
                    \left(
                        \frac{1-\hat{\eta}}{\hat{\eta}}
                    \right)
                \right\rangle_\refDensity
                -
                \left\langle
                    \frac
                    {\hat{\eta}\circ \vmap}
                    {1-\hat{\eta}\circ \vmap}
                \right\rangle_{\preDensityParam}
            \right\}
        \end{equation}
        \begin{equation}\label{relentderiv}
            \frac{
                \md
            }
            {
                \md \cgParam_i
            }
            \Qual^{\text{RE}}
            \left[
                \preDensityParam,\refDensity; \vmap
            \right]
            =
            - \beta
            \left\langle 
                \frac{\hat{\eta}^{\dagger}\circ \vmap}
                {1-\hat{\eta}^{\dagger}\circ \vmap}
            \right\rangle_{\preDensityParam}
            \left\langle
            \frac{
                \partial \prePot_\cgParam
            }
            {
                \partial \cgParam_i
            }
            \right\rangle_{\preDensityParam}
            + \beta
            \left\langle
                \left(
                    \frac{\hat{\eta}^{\dagger}\circ \vmap}
                    {1-\hat{\eta}^{\dagger}\circ \vmap}
                \right)
            \frac{
                \partial \prePotParam
            }
            {
                \partial \cgParam_i
            }
            \right\rangle_{\preDensityParam}
        \end{equation}.
    \end{widetext}
    This comprises a full residual and associated gradient for optimization. 
    However, in practice, the
    loss functions are poorly behaved: pointwise values of $\hat{\eta}=1$
    easily create a divergent residual value (identical to the corresponding
    situation with the traditional relative entropy estimation methods).
    Fortunately, the optimal $\eta$ is shared among all proper
    losses.\cite{reid2011} As
    a result, $\hat{\eta}^\dagger$ can be similarly discovered with the corresponding
    statement using the log-loss\cite{james2013introduction,reid2011}
    \begin{equation}\label{loglearn}
        \hat{\eta}^\dagger =
        \argmin_{\hat{\eta}}
        \left\{
            \left\langle
                \log \hat{\eta}
            \right\rangle_\refDensity
            +
            \left\langle
            \log (1 - \hat{\eta} \circ \vmap)
                \right\rangle_{\preDensityParam}
            \right\}
    \end{equation}
    while the gradient estimation remains unchanged. To summarize, the models
    trained in this article indirectly minimize Eq. \eqref{relentlearn} by producing
    derivatives over $\cgParam$ via Eq. \eqref{loglearn} and Eq. \eqref{relentderiv},
    where $\hat{\eta}^\dagger$ retains the same meaning across equations. This
    equality only holds assuming that $\hat{\eta}^\dagger =
    \eta$; incomplete
    $\fpairspace$ can cause the resulting $\hat{\eta}^\dagger$'s to differ.

    The numerical examples section \ref{results_p} are computed in the following
    way. First, the CG (or pre-CG) model is represented using a 
    molecular force-field and samples are generated using standard molecular
    dynamics software. These samples are mapped if necessary using $\vmap$.
    Reference examples are similarly generated and mapped using $\map$ if
    needed. The variational estimator is represented using either a neural
    network or through logistic regression, which implement Eq. \eqref{loglearn}.
    The estimator then is trained on the reference and model samples. Finally, the
    gradient is calculated using the output of the variational estimator, Eq.
    \eqref{relentderiv}, and the model samples; this gradient is then used to
    update the model parameters. This process is iterated, although the
    reference sample is not regenerated. The variational estimators are not fed
    the Cartesian coordinates of the input system directly; instead, various
    features are calculated for each frame, and these features are given as
    input to the variational estimator. This has the effect of constraining that
    $\hat{\eta}$ be a function of these features. Additional points on each of
    these details is discussed for each example or may be found in the appendix.

    In some cases of ARCG, including the case of $f$-divergence estimation, the
    functions achieving the inner maximum with an complete $\fpairspace$ can be
    expressed as a pointwise functions of the mapped distributions.
    Specifically, as noted in Eq. \eqref{etadef} the optimal witness function $\eta$
    in the case of relative entropy is expressible as a function of the
    conditional class densities ($\cgDensity$ and $\refDensity$).  This can guide how elements of a tractable
    $\fpairspace$ are parameterized. When the algebraic forms of $\refDensity$
    and $\cgDensity$ are known to be functions of summary statistics of their
    respective systems (e.g., the inverse 6 and 12 moments in a traditional
    Lennard-Jones potential\cite{lennard1931cohesion}), we can often express an
    complete $\fpairspace$ exactly with a manageable number of terms;
    however, this is not true of practical bottom-up CG application: the form of
    the mapping operator does not provide us with an algebraic understanding the
    implied mapped free energy surfaces. However, the resulting $\eta$ does
    share invariances with the free energy surfaces it is composed of (e.g.,
    rotational and translational invariances). 

    The integrals characterizing the variational residual are computationally
    approximated as sample averages. Optimizing a function using a sample
    average introduces the possibility that the function which maximizes the
    sample average is a poor approximation of the function which maximizes
    population average. In the context of classification this error is captured by
    considering whether the classifier is overfitting the data sample.  There
    are multiple strategies to overcome this;\cite{james2013introduction} in the
    current study we use $l2$ regularization and only allow the variational
    estimator to update a limited number of times at each iteration. When optimizing
    examples with flexible potentials and large feature sets (for example, the
    water and methanol models presented in the next section), we have found that
    using a neural network quickly overfits the data provided, even with
    relatively strong regularization. However, reducing the number of iterations
    allowed at each step of variational optimization causes 
    the neural network to exhibit considerable hysteresis between iterations,
    causing the force-field being optimized to orbit around an ideal solution. To
    ameliorate this we use logistic regression in these more complex cases,
    where the solution to the logistic regression is optimized using a limited
    number of iterations of l-BFGS. Note that the output of logistic regression
    readily affords estimates of the class conditional probabilities, which in
    turn directly connects its optimal solution to $\eta$. 

    The issues associated with overfitting and hysteresis are intrinsically
    connected the size of the finite samples used to approximate the integrals
    in Eq. \eqref{icgresidual}. Overfitting would be reduced by increasing the
    sample size, which in turn would allow additional optimization at each
    variational iteration and would therefore reduce hysteresis. In practice, we
    have found that increasing the sample size to the point that the hysteresis
    is removed slows down the rate of force-field optimization considerably due
    to the time needed to both generate molecular samples and evaluate high
    dimensional gradients. This difficulty naturally suggests the use of
    modified sampling strategies to reduce the discrepancy between the sample
    and population averages. As Eq. \eqref{icgresidual} only involves
    expectation values any modified sampling scheme which allows for the
    calculation of an unbiased ensemble average is a candidate for this
    strategy. It is worth noting that ARCG selects an optimal observable
    function based on the ensemble averages of a large set of candidate
    observables, and that the error for each observable may be different,
    complicating the use of variance reduction techniques which are designed for
    a single observable. It would also be possible to improve the sampling of
    the feature space on top of which $\fpairspace$ is represented; for
    example, if $\fpairspace$ is represented by a neural
    network acting on two
    statistics calculated for each sample, free energy estimation may be used to
    better resolve the joint distribution of these two statistics themselves. This
    approach could be extended to produce an parameterization method which
    improves the observable estimates on the fly, such as
    that in \citet{abrams2012fly}. While we
    have not pursued these strategies here, future applications to more complex
    systems will likely need consider these options.

    The variational search over possible $\hat{\eta}$ was either performed via a
    neural network outputting class probability predictions penalized via the
    log-loss or through logistic regression. Logistic regression
    was used in the cases of examples using $b$-spline based
    potentials and neural networks were
    used in all other cases. All neural
    networks used in examples in this paper utilized a simple feed-forward
    architecture with at least two layers (not including the input and output layer). 
    The results were found to be insentive to the architecture chosen, and the
    specific architectures used may be found in appendix \ref{num_details}.
    All internal nodes used rectified linear activation functions with
    the output normalized via softmax. The duality with classification
    underpins the utility such traditional choices have in our
    variational search.

    In practice, we have noticed that ARCG optimization may suffer from
    instability, especially when optimizing the parameters of a model that
    produces a distribution significantly different than its optimization
    target. This issue can be noted by observing that the classifier achieves
    100\% accuracy during parameterization, producing uninformative gradients.
    In these cases we find that an effective strategy is to introduce standard
    Gaussian noise into both the model and reference samples; the variance of
    this noise is gradually reduced to zero as the optimization progresses. It
    is likely that a correct local minima is achieved in this case as the
    optimization appears stationary at the end of
    minimzation, but it is
    unclear if the selection of a specific local minima is biased using this
    strategy.

    A public proof-of-concept python/Lammps based implementation is available at
    the weblink https://github.com/uchicago-voth/ARCG. This code base makes
    extensive use of the theano, theanets, pyLammps, numpy,
    scikit, and dill libraries.
    All computational examples presented in this paper may be found in the test
    portion of this code, which includes the complete settings used to generate
    the data used.  Visualizations and analysis were performed with the matplotlib and
    seaborn libraries, as well as the base plotting system,
    rgl, and data.table packages in R.
    Extensions providing scalability for more complex systems and potentials
    will be considered in future work.

\section{Results} \label{results_p}

    The relative entropy approach described in section \ref{impl_p} was applied
    to five test systems. First, a simple single component 12-6 Lennard-Jones
    (LJ) system was optimized to approximate a reference LJ system at the same
    resolution (no virtual particles were present, and no coarse-graining of
    either the reference or model was performed).  Second, a system representing
    two bonded real particles where force is partially mediated by a single
    harmonically bonded virtual particle was optimized to approximate a
    reference system of the same type. Third, a binary LJ liquid undergoing
    phase separation was simulated and optimized after particles of a single
    type had been integrated out; this distribution was fit to match a similarly
    integrated binary LJ system. Fourth, a CG model using pairwise $b$-spline
    interactions and a single site per molecule was used to approximate liquid
    methanol. Fifth, a CG model using pairwise $b$-spline interactions and a
    single site per molecule was used to approximate liquid water.  In these
    cases we observed good convergence of suitable correlation functions;
    however, in cases with virtual particles we found that numerically
    recovering the known parameters of the reference system is difficult; in
    other words, it seems likely that the parameter space is either redundant or
    sloppy,\cite{transtrum2015perspective} with similar correlation functions
    arising from distinct parameter sets.  We note that while the potentials
    considered here are relatively simple, ARCG is fully applicable to more
    complex potentials such as those in \citet{zhang2018deepcg}.

    The first three examples considered here are theoretically able to capture
    the reference distributions used for fitting (i.e., the model optimized is
    not misspecified). This is ensured by generating reference data using a
    force-field that is directly representable by the CG force-field family. For
    example, the LJ CG model in the first example was optimized to reproduce the
    statistics generated by a particular LJ reference potential.  Additionally,
    when either the reference or model are modified using a mapping function,
    this mapping operator is forced to be the same between the two systems, and
    the reference data is again produced using a force-field which is
    expressible by the CG model.  For example, in the case of the virtual
    solvent LJ system, a distinct system of binary LJ particles was simulated for both
    the model and reference data samples, each with differing parameter sets.
    Both systems then had the particles of a specific shared type integrated
    out. The resulting integrated distributions were then the basis of
    comparison used to train the model parameters (with new statistics being
    created for the CG model at each iteration). This is not true for the
    examples approximating water and methanol: here, the CG model is
    approximating the mapped distributions using a pairwise potential and is
    unable to capture the true free energy surface.

\subsection{Lennard-Jones Fluid}

\begin{figure}[h]
        \centering
        \includegraphics[width=8cm]{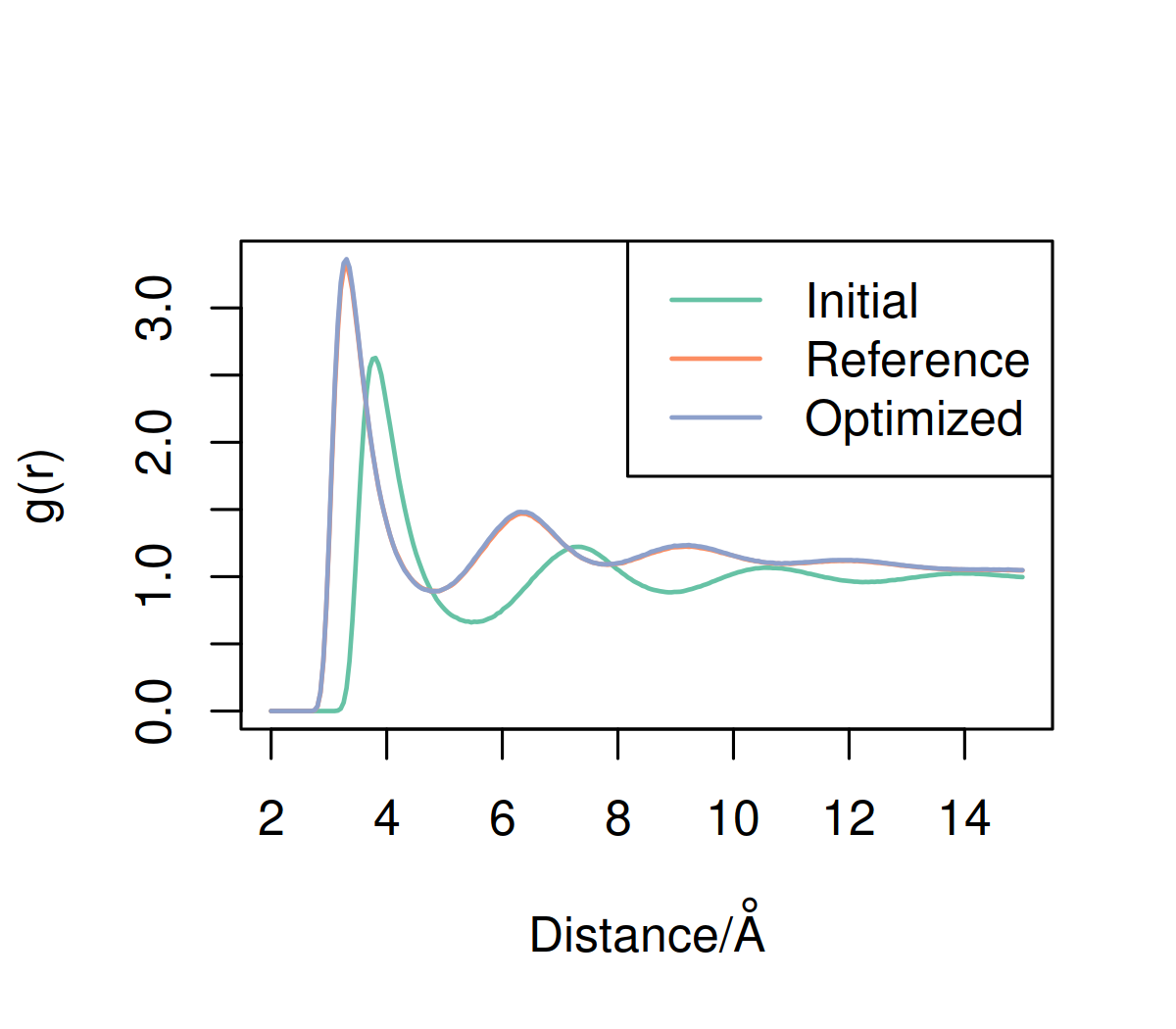}
        \caption{Radial distribution functions calculated for 
                 the unoptimized system $A_{\text{initial}}$, the reference system 
                 $B$, and the optimized system $A_{\text{opt}}$.}
        \label{fig:lj_rdf}
    \end{figure}

    A single component 12-6 LJ fluid was simulated with 864
    particles at 300K (the potential form  is given in Eq. \eqref{ljpot} 
    with $r_{ij}$ denoting the
    Euclidean distance between particles $i$ and $j$).
    \begin{equation}\label{ljpot}
        U(\pos^{3\cgnp}) = 4 \epsilon  \sum_{i>j} \left[ 
                \left(\frac{\sigma}{r_{ij}}\right)^{12}
                -
                \left(\frac{\sigma}{r_{ij}}\right)^6
                \right]
    \end{equation}
    The system was simulated at constant NVT conditions using a 
    Langevin thermostat with
    coupling parameter set to 100.0 fs and a timestep of 1.0 fs.  No virtual
    particles were present; i.e., $\vmap$ and $\map$ are set to be the identity
    function. Inverse sixth and twelfth moments were used as input to the
    variational estimator (in this case, this set of features is known to be
    complete, see appendix \ref{num_details} for details). System $A_{\text{initial}}$ with
    $\epsilon_{A_{\text{initial}}}=0.6 \text{kcal}/\text{mol}$ and 
    $\sigma_{A_{\text{initial}}}=3.5 \angstrom$ was optimized to match the
    statistics of system $B$ characterized by $\epsilon_B=0.75
    \text{kcal}/\text{mol}$ and $\sigma_B=3.0 \angstrom$. Upon
    optimization, the parameters of $A$ were seen to quantitative converge to
    those of $B$: $\epsilon_{A_\text{opt}}=0.746
    \text{kcal}/\text{mol}$
    and $\sigma_{A_\text{opt}}=3.00 \angstrom$. 
    Additionally, convergence of the
    pairwise correlation functions (fig. \ref{fig:lj_rdf}) was observed.
    The initial parameters of $A$
    resulted in a homogeneous liquid, while those of system $B$ (and system $A$
    upon optimization) resulted in liquid-vapor coexistence.   During
    training Gaussian noise was used to smooth out initial gradients to resolve
    initial soft wall differences; this noise is reduced to zero by the end of
    optimization.  Optimization was performed using
    RMSprop\cite{hinton2012neural} with individual
    rates for each parameter. These results demonstrate good
    convergence properties with small parameter sets when no virtual particles
    are considered in the pre-CG resolution. 

\subsection{Virtual Bond Site}

\begin{figure}[h]
        \centering
        \includegraphics[width=8cm]{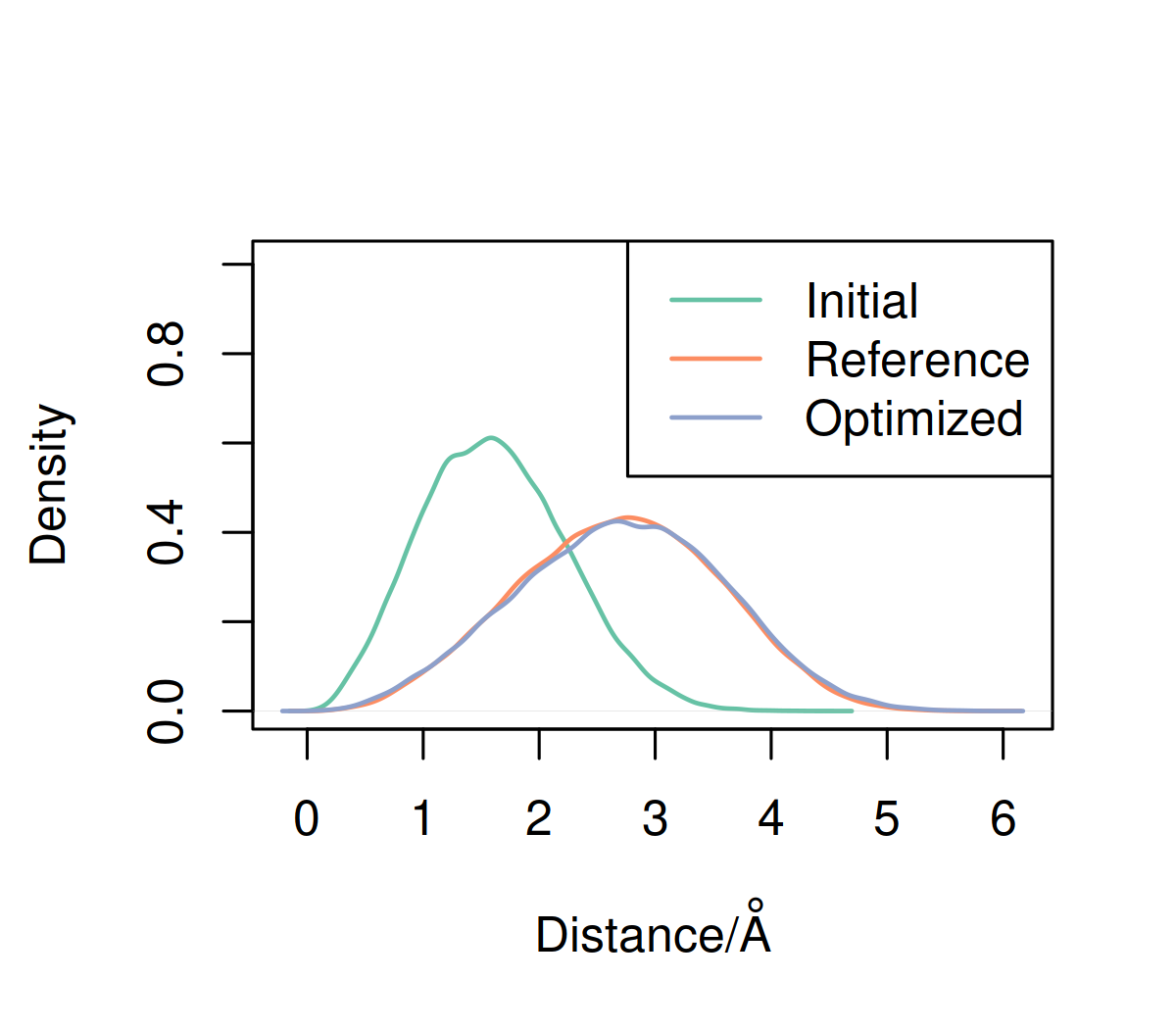}
        \caption{Bond distance distribution functions calculated for 
                 the unoptimized 
                 system $A_{\text{initial}}$, the reference system $B$, and the 
                 optimized system $A_{\text{opt}}$.}
        \label{fig:vsitedis}
    \end{figure}

    A system of three particles completely connected via
    harmonic bonds was simulated at 300 K. 
    The system was propagated in constant NVT conditions using a Langevin
    thermostat with coupling parameter set to 100.0 fs and a
    timestep of 1.0 fs.
    Two types of particles are present; we denote the types of 
    the particles $X,Y,X$. 
    Upon application of $\map$ and $\vmap$, the $Y$ particle is removed,
    resulting in a system composed of two particles of type
    $X$ (i.e., the $Y$ particle is a virtual site). This mapped
    system is optimized using the distance between the two $X$ particles as input to the
    discriminator; in this case, this feature set is complete. Initial,
    optimized, and reference parameters are seen in table \ref{table:vsite}.
    \begin{table}[h]
        \centering
        \renewcommand*{\arraystretch}{1.3}
        \begin{tabular}{| l | c | c | c | c |}
            \hline
            System & $x_{XY}/\angstrom$ & $k_{XY} /
            \frac{\text{kcal}}{\text{mol}}\angstrom^{-2}$ & 
            $x_{XX}/\angstrom$ & $k_{XX} /
            \frac{\text{kcal}}{\text{mol}}\angstrom^{-2}$ \\ \hline
            $B$ & 2 & 2.7 & 2.3 & 0.4 \\ \hline
            $A_\text{initial}$ & 0.65 & 2.2 & 1.4 & 0.15 \\ \hline
            $A_\text{opt}$ & 1.70 & 2.06 & 2.66 & 0.224 \\ \hline
        \end{tabular}
        \caption{Parameters for systems with virtual bonded sites. $x$ denotes
        the zero energy point of the bond while $k$ denotes bond strength.
        Subscripts specify the particle types between which the bond acts.
        System $A_\text{initial}$ was optimized to match system $B$, resulting
        in $A_\text{opt}$.}
        \label{table:vsite}
    \end{table}
    Optimization was performed using RMSprop.
    Convergence to a specific parameter set that reproduces
    observed correlations (fig. \ref{fig:vsitedis}) is fast; however these
    parameters differ from the parameters of the reference system. Additional
    simulations were run where the CG model was initialized with 
    parameters set to those of the reference system (results not shown); in this case, we
    observed local diffusion around a small set of parameters including the true
    set. This suggests that virtual particles may create degeneracy in model
    specification in practice (i.e., even if the model parameters are
    identifiable, the specification is sloppy).
    This case represents an application where a pairwise force-field may be
    augmented via bonded virtual particles to create modified correlations.  For
    example, a heterogeneous elastic network\cite{lyman2008systematic} 
    may be augmented by introducing virtual particles to facilitate 
    higher order correlations.

\subsection{Virtual Solvent Lennard-Jones Fluid}

\begin{figure}[h]
        \centering
        \includegraphics[width=5cm]{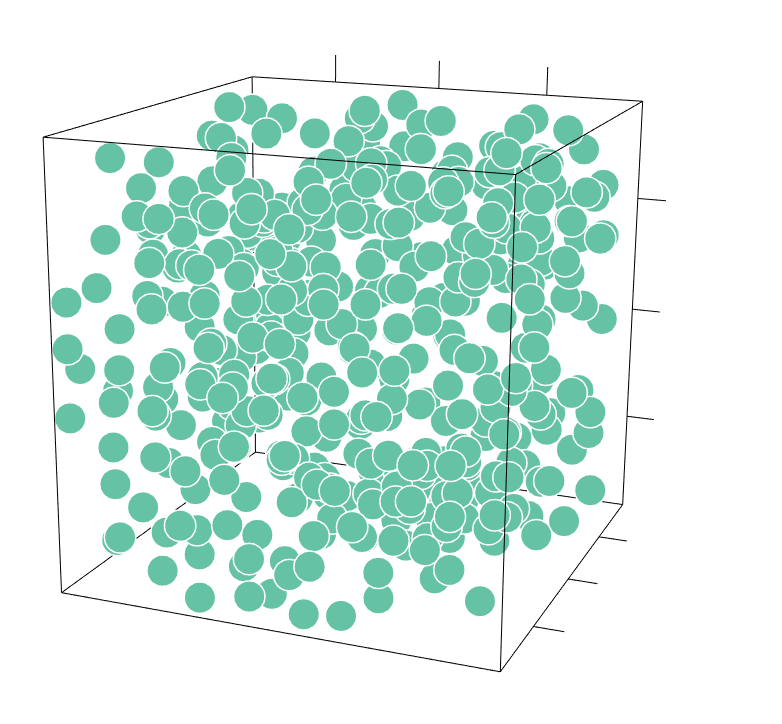}
        \includegraphics[width=5cm]{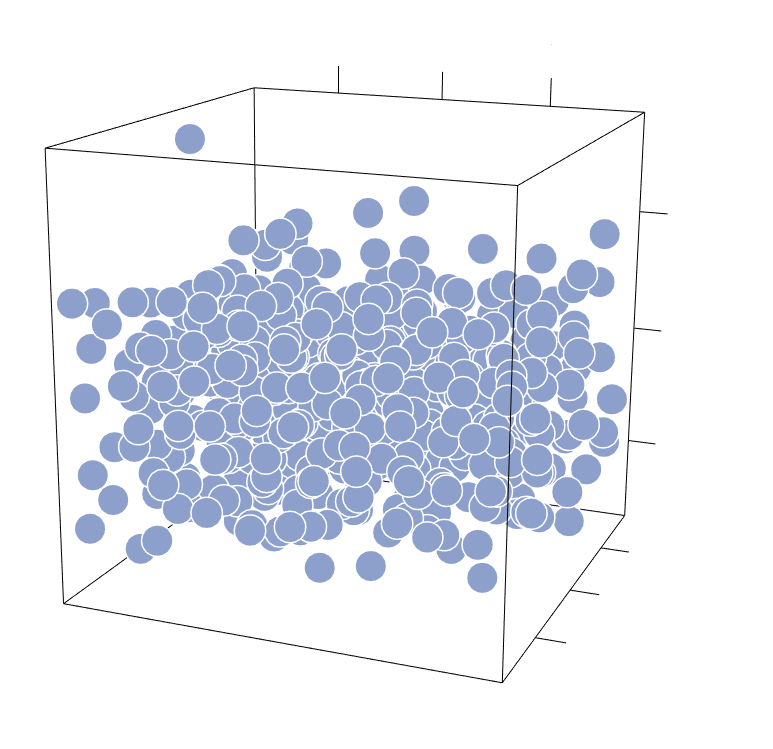}
        \includegraphics[width=5cm]{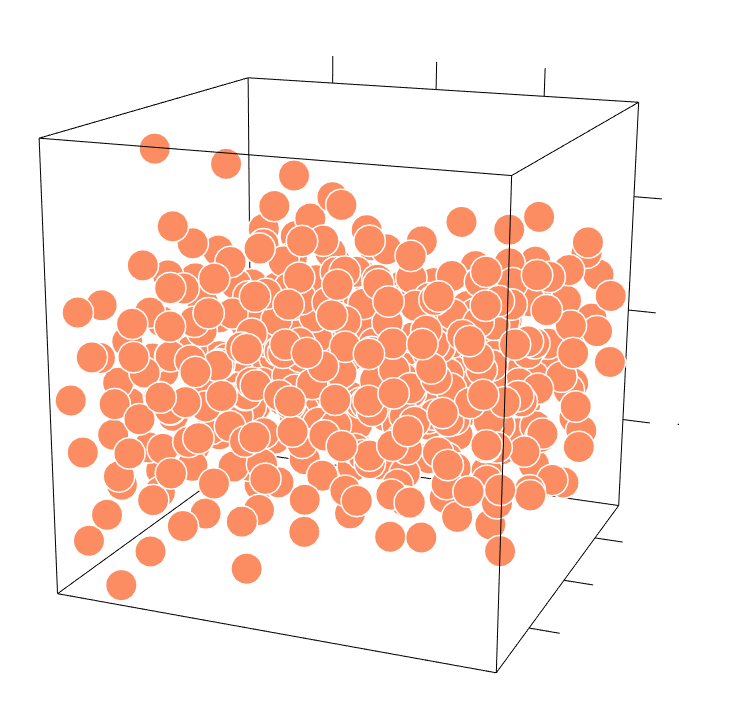}
        \caption{Sample configurations of the unoptimized model (green), the optimized
        model (blue) and the reference data (orange). Configurations are shown at
        the resolution of comparison, i.e., after the application of $\map$ and
        $\vmap$. Slab type formation, similar to that present in the
        optimized model, is seen after parameter optimization.}
        \label{fig:vpconfig}
    \end{figure}

\begin{figure}[h]
        \centering
        \includegraphics[width=8cm]{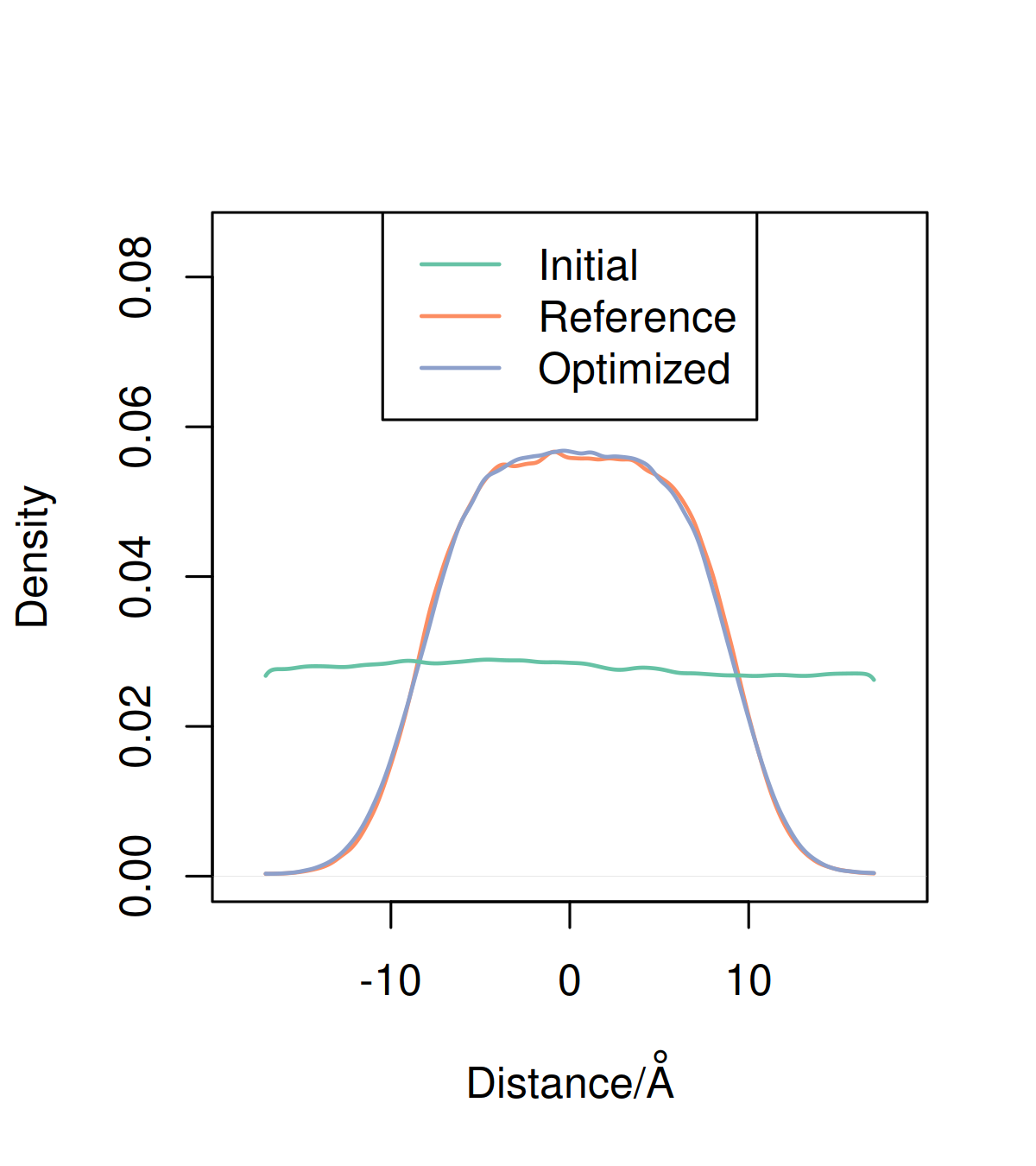}
        \caption{Probability densities across the slab type
        formations present in the integrated binary LJ systems 
        (along the $z$ axis of the simulation box).
        No slab structure is present in the initial model.}
        \label{fig:vpcorr}
    \end{figure}

    A binary system composed of 864 LJ particles of types $X$ and $Y$ was
    simulated at 300 K.  The system was simulated at constant NVT conditions
    using a Langevin thermostat with coupling parameter set to 100.0 fs and a
    timestep of 1.0 fs.  Equal numbers of $X$ and $Y$ particles were present
    prior to the application of mapping operators; upon application all
    particles of type $Y$ were removed (i.e., the $Y$ particles are virtual
    sites).  The target system was parameterized to undergo phase coexistence,
    while the unoptimized CG model was well mixed.  Parameters are found in
    table \ref{table:blj}. Optimization was performed using RMSprop with rates
    adjusted for each parameter. Gaussian noise was used to stabilize initial
    training. Visual inspection of representative molecular configurations
    showed greatly improved similarity for the optimized parameter set (fig.
    \ref{fig:vpconfig}).  Again, while convergence of correlation functions is
    readily observed (fig. \ref{fig:vpcorr}), parameters do not converge to
    those of the reference system, likely due to sloppiness in specification.

    \begin{table}[h]
        \centering
        \renewcommand*{\arraystretch}{1.3}
        \begin{tabular}{| l | c | c | c | c | c | c |}
            \hline
            System & $\sigma_{XX} / \angstrom$ & $\epsilon_{XX}
            / \frac{\text{kcal}}{\text{mol}}$ & 
            $\sigma_{YY} / \angstrom$ & 
            $\epsilon_{YY} / \frac{\text{kcal}}{\text{mol}}$ & 
            $\sigma_{XY} / \angstrom$ & 
            $\epsilon_{XY} / \frac{\text{kcal}}{\text{mol}}$ \\ \hline
            $B$ & 0.7 & 3.6 & 0.7 & 3.6 & 0.35 & 3.5  \\ \hline
            $A_\text{initial}$ & 0.6 & 3.5 & 0.6 & 3.2 & 0.5 & 3.1 \\ \hline
            $A_\text{opt}$ & 0.713 & 3.600 & 0.722 & 3.594 & 0.349 & 3.494 \\ \hline
        \end{tabular}
        \caption{Parameters for systems with virtual bonded sites. $x$ denotes
        the zero energy point of the bond while $k$ denotes bond strength.
        Subscripts specify the particle types between which the bond acts.
        System $A_\text{initial}$ was optimized to match system $B$, resulting
        in $A_\text{opt}$.}
        \label{table:blj}
    \end{table}
    This case is representative of the situation where higher order correlations may
    be captured by the addition of virtual solvent particles. For example, the
    hydrophobic driving force underlying a CG lipid bilayer could be facilitated
    by a virtual solvent. This is distinct from using traditional explicit
    solvent where each solvent molecule is directly connected to the FG
    reference system: there, the behavior of the solvent is incorporated into
    the quality of the model, as where the approach of ARCG ignores the direct
    solvent behavior.

\subsection{Single Site Methanol}

\begin{figure}[h]
        \centering
        \includegraphics[width=8cm]{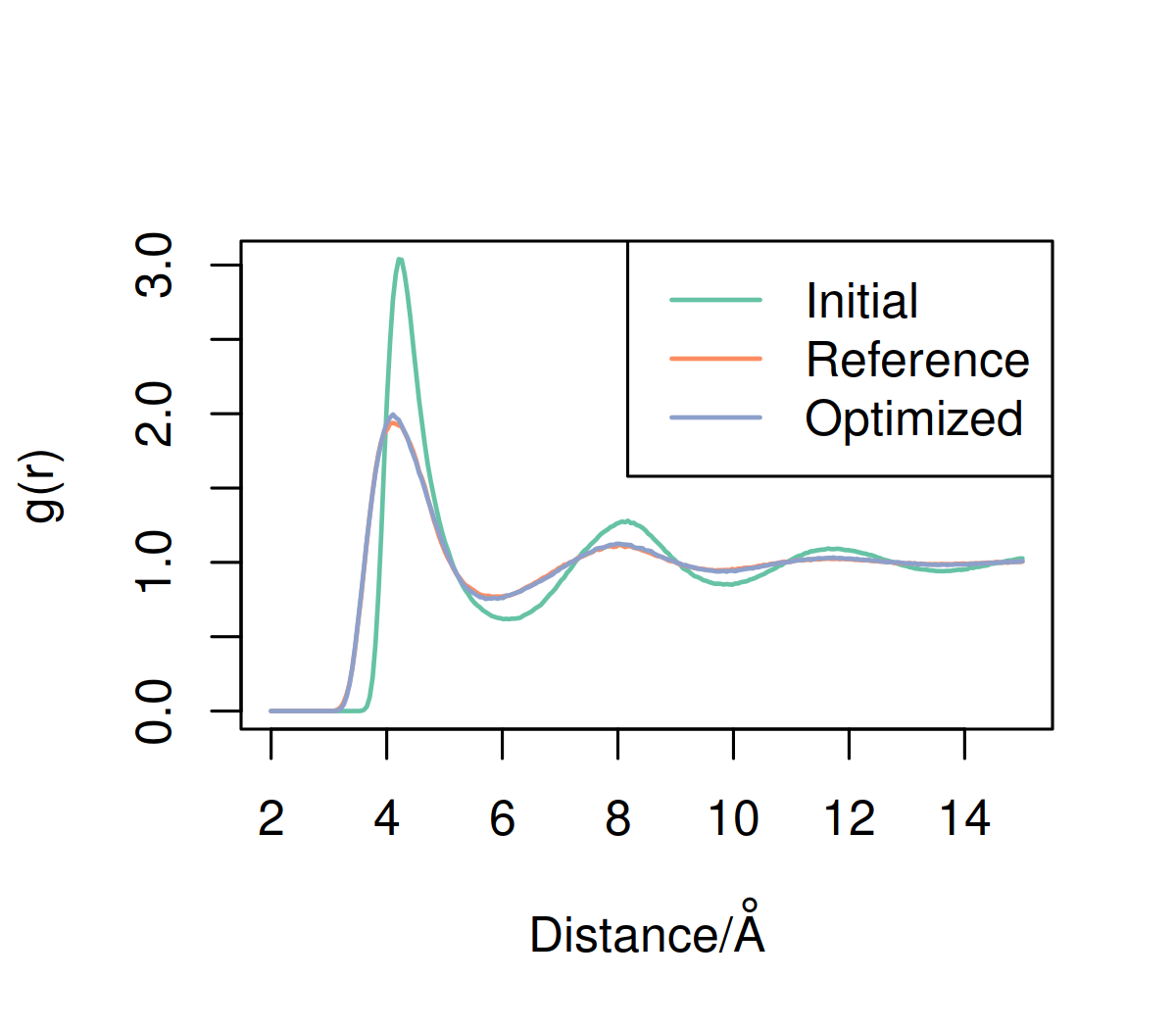}
        \caption{Radial distribution functions for the reference, initial, and
        optimized methanol systems. Note that the optimized and reference RDFs are
        nearly within line thickness.}
        \label{fig:meth_rdfs}
    \end{figure}

\begin{figure}[h]
        \centering
        \includegraphics[width=8cm]{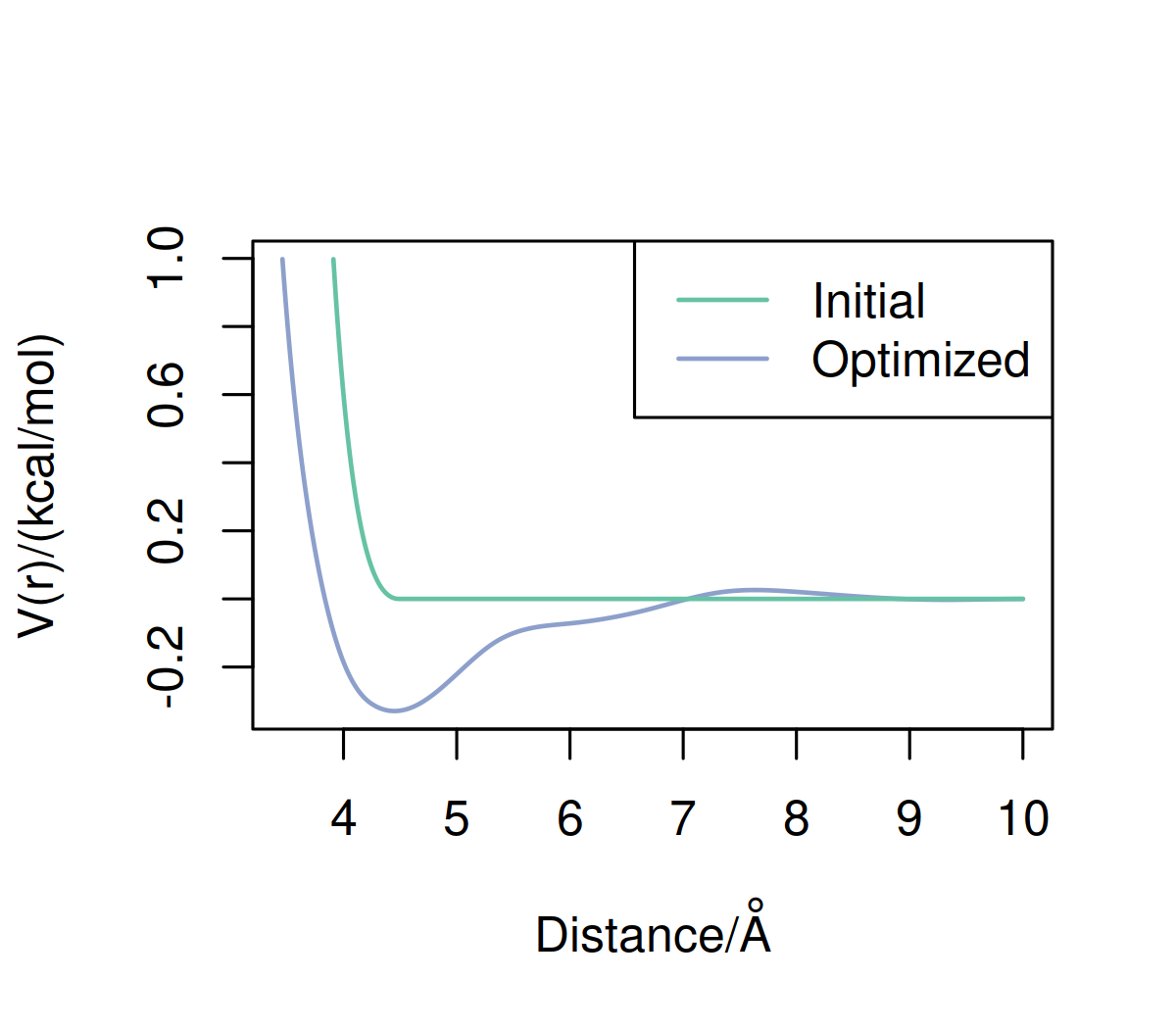}
        \caption{Pairwise potential functions characterizing the initial and
        optimized methanol systems.}
        \label{fig:meth_pots}
    \end{figure}

    Methanol was modeled using single site CG liquid. The reference atomistic
    (FG) trajectory of 512 molecules was simulated in the NVT ensemble at 300K
    with a Nose-Hoover damping time of 1 ps after NPT
    equilibration at 1 atm. The
    OPLS-AA\cite{jorgensen2005potential,dodda20171,dodda2017ligpargen} forcefield was used in
    the atomistic system.
    The FG system was mapped to the to the CG resolution by retaining only the
    central carbon; no virtual sites were present in the CG system. The CG
    potential was described using a pairwise $b$-spline potential using 15
    equally spaced knots
    and a 10 $\angstrom$ cutoff (the last three control points were set to zero to
    enforce a smooth decay at the cutoff). The CG system was run at the same
    temperature and volume as the FG system using a Langevin coupling parameter
    of 100 fs and a timestep of 1 fs. The starting potential used for the simulation was a WCA potential
    (fig \ref{fig:meth_pots}). Quantitative reproduction of the radial distribution function
    was observed (fig \ref{fig:meth_rdfs}).  Convergence was smoothing using Gaussian noise
    whose standard deviation decayed to zero by the end of the optimization.

\subsection{Single Site Water}

\begin{figure}[h]
        \centering
        \includegraphics[width=8cm]{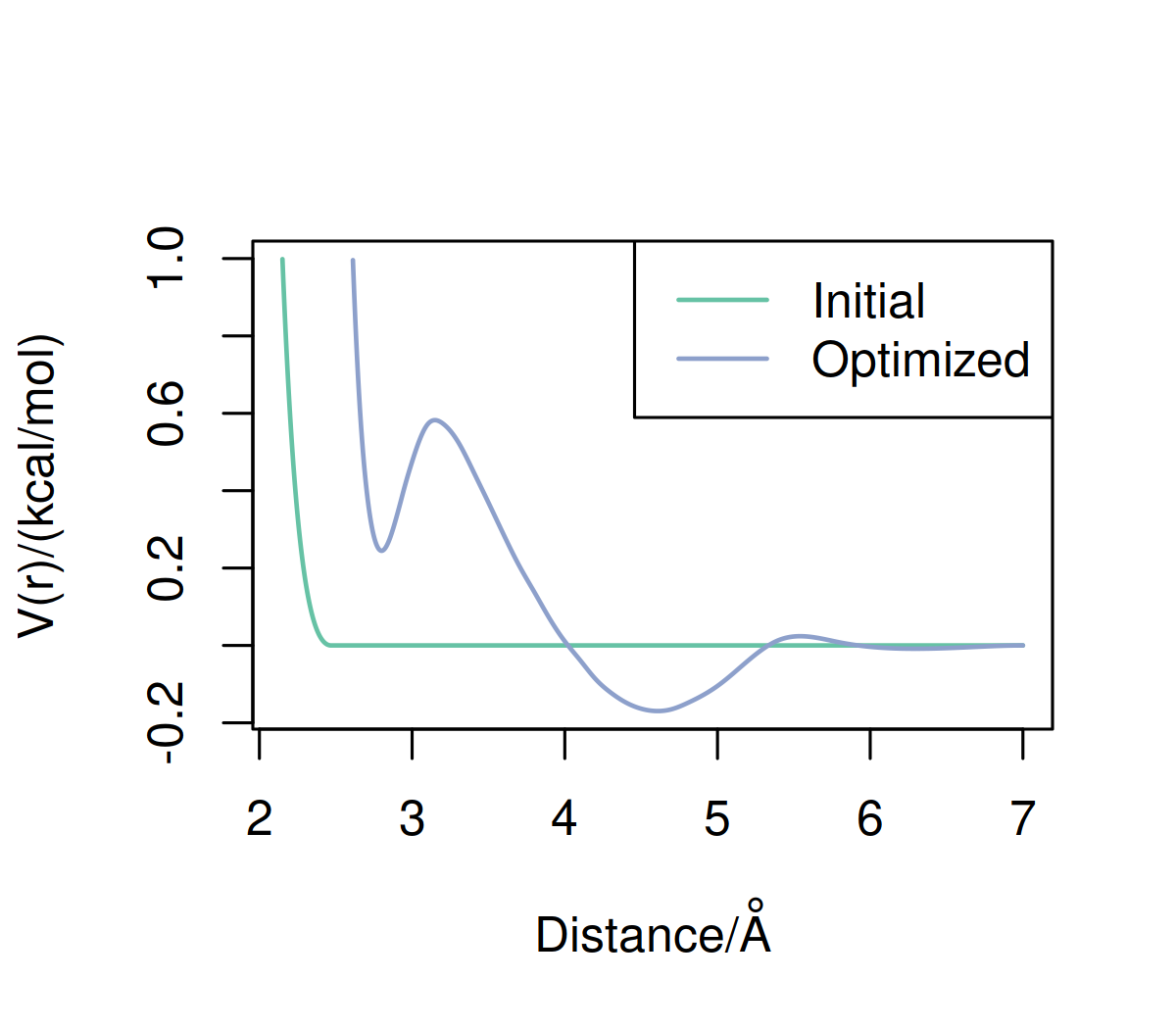}
        \caption{Pairwise potential functions characterizing the initial and
        optimized water systems.}
        \label{fig:water_pots}
    \end{figure}

\begin{figure}[h]
        \centering
        \includegraphics[width=8cm]{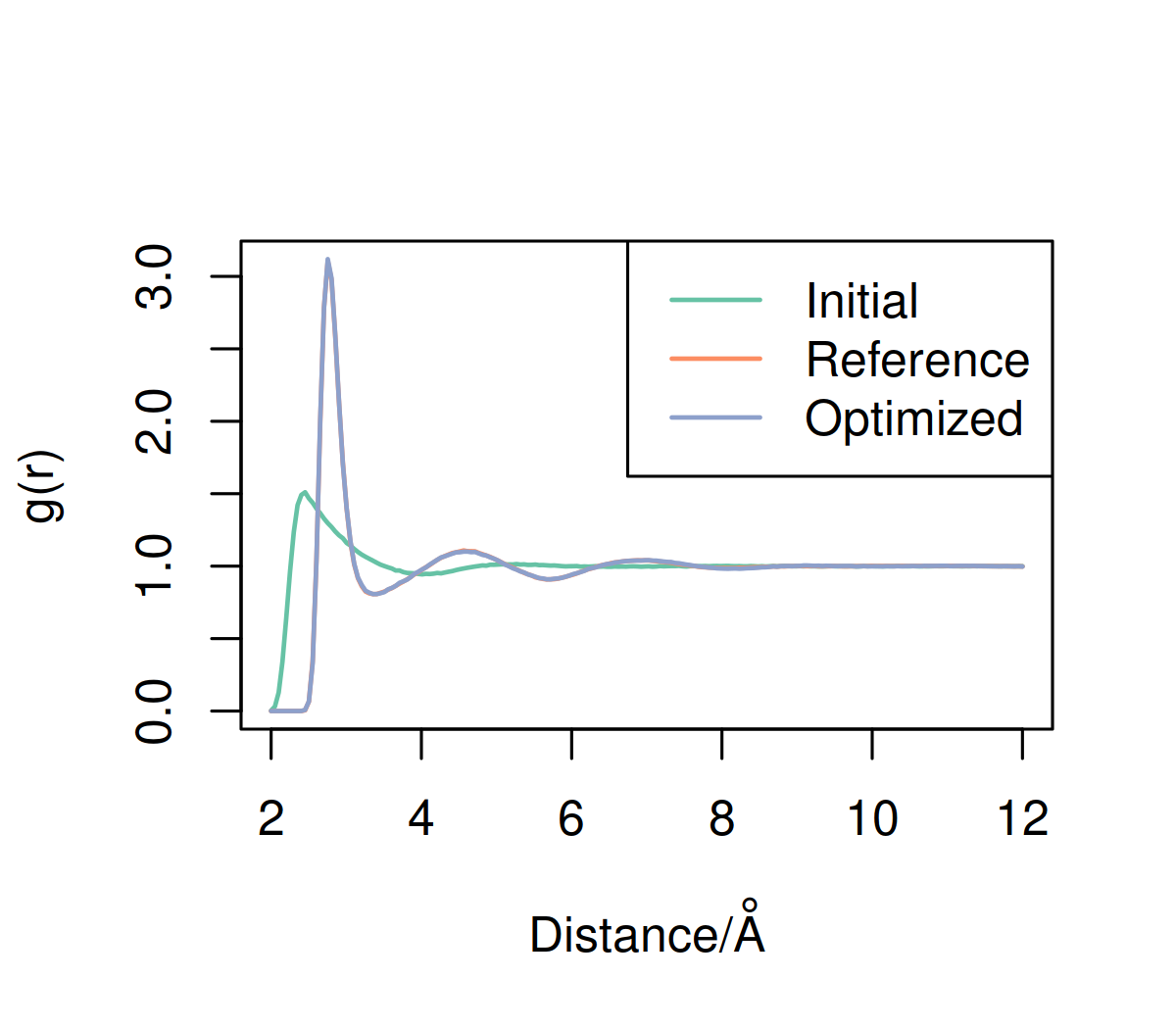}
        \caption{Radial distribution functions for the reference, initial, and
        optimized water systems. Note that the optimized and reference RDFs are
        within line thickness.}
        \label{fig:water_rdfs}
    \end{figure}

    A single site model of water was trained using 512 molecules of SPC/E water
    simulated at 300K. The molecular (FG) system was equilibrated at 1 atm and
    production NVT samples were produced with the Nose-Hoover thermostat with a 1 ps
    damping time.  The mapping connecting the FG system to the CG system was the
    center of mass mapping; no virtual sites were present in the CG system. The
    CG system potential was limited to pairwise interactions with a 7 $\angstrom$ cutoff
    and was parameterized using $b$-splines with 37 knots (see
    appendix \ref{num_details} for knot locations and
    further details). The last three spline control points
    were set to zero to enforce continuity at the cutoff. The starting potential
    used for the simulation was a WCA potential (fig
    \ref{fig:water_pots}). CG simulations were run at the
    same volume as the FG system with a Langevin thermostat,
    whose coupling parameter was set to 100 fs, and a 1 fs
    timestep. Quantitative reproduction of
    the radial distribution function was obtained (fig
    \ref{fig:water_rdfs}). Gaussian noise was
    used initially to smooth convergence and was tapered to a standard deviation
    of zero by the end of the optimization.

\section{Discussion} \label{disc_p}

    In previous sections we have described a broad new class of variational
    statements for optimizing CG models and described methods for their
    optimization by utilizing the theory underpinning adversarial models in ML.
    Subsequently we have shown that it is possible to parameterize a CG model
    via ARCG at a coarser resolution than that native to the CG Hamiltonian.  A
    clear application of ARCG is the parameterization of
    models that contain
    virtual sites; however, the CG distribution may be critiqued at any coarser
    resolution, providing the intriguing ability to control what aspects of a CG
    model are visible for optimization purposes. In the process of doing so we
    showed that gradients needed at each step of divergence minimization can be
    reformulated as modifying the system Hamiltonian to minimize the value of a
    specific observable, but that this observable depends on the 
    distributions being considered at that step of optimization.  We note that
    more generally the method presented can be used to calculate the KL
    divergence (and any of the other divergences discussed) between
    distributions for which no probability density/mass is known and for which
    one cannot be approximated via kernel density approximation or binning.

    Beyond our central results we have provided work and discussion on two 
    supporting topics.
    \begin{enumerate}
        \item

            We have provided comparisons to multiple contemporary methods
            for CG parameterization. In certain cases we have shown that
            divergences characterizing their configurational variational
            principles can be used in ARCG modeling. In one case we showed that
            classifier based approaches bear striking but not complete
            similarity to the presented approach. In the remaining cases we
            have discussed how decoupling the resolution at which we critique
            a model from the resolution of the CG Hamiltonian creates
            difficulties in said approaches.

        \item

            We have provided a set of sufficient conditions for momentum
            consistency in the case of virtual sites and described how these
            conditions may be extended. These are closely related to consistency
            requirements for traditional bottom-up CG models.

    \end{enumerate}

    Additionally, we have provided simple numerical examples (and a public
    computational implementation) for which we have
    optimized CG potentials to match specific distributions, some of which utilize
    CG virtual particles. The results show quantitative agreement for calculated
    correlations, visual agreement, and qualitative agreement in
    matching exact coefficients when the answer is known
    (quantitative agreement is seen when virtual particles
    are not present). Difficulties in
    convergence appear to be either due to instability in the parameterization process
    or sloppiness in the model specifications.  The manner in which
    this will affect realistic systems is yet to be seen, but may present a
    significant challenge. It is clear that in the most general case parameter uniqueness
    is not guaranteed: if CG consistency can be obtained without virtual
    particles, then a model that can both decouple the virtual particle interaction
    from the real particles and modify the behavior of the virtual particles
    independently of said coupling will inherently be nonidentifiable. 
    Additionally, it is likely that in the case of $f$-divergence based ARCG
    optimization that a relatively good initial hypothesis for the CG potential
    may be necessary, or significant amounts of noise must be added initially
    during optimization.

    There are multiple additional studies that could naturally expand and
    clarify the results presented.
    \begin{enumerate}
        \item

            The methods provided can be applied to approximate nontrivial
            molecular systems without virtual particles. This will require
            multiple steps: first, the proof-of-concept software framework
            presented will have to be expanded for larger system sizes. Second,
            the training method used will have to be developed such that it
            remains stable, whether through the systematic addition of noise or
            the use of enhanced sampling techniques. Third, the feature-space
            used to index $\fpairspace$ will likely have to be correctly
            engineered based on knowledge of the FG and CG Hamiltonians. All
            three of these are tractable challenges.

        \item

            The effect of using virtual particles should be investigated both
            computationally and theoretically, as previous analysis on
            incomplete basis sets (e.g., that on relative entropy and MS-CG
            \cite{rudzinski2011coarse}) does not apply transparently.  In the
            process of doing so a better theoretical understanding of how to
            utilize these methods to capture specific higher order correlations
            in the training data should additionally be investigated, possibly
            leading to new ways in which bottom-up CG parameterization may be
            tuned to reproduce specific novel correlation functions.

        \item

            The effect of various divergences on training approximate CG models
            should be investigated theoretically and through simulation. This
            will facilitate the design of CG
            parameterization methods that have
            different biases in the approximations they produce when coupled
            with realistic CG potentials. This applies to not only to various
            $f$-divergences but also the wider set of divergences not heavily
            discussed in this article, such as the
            Wasserstein,\cite{arjovsky2017wasserstein}
            Sobolev,\cite{mroueh2017sobolev} Energy,\cite{2016arXiv160602556B} and
            MMD\cite{dziugaite2015training}
            distances. The Wasserstein and Energy distances share the
            interesting property of taking into account the spatial organization
            of the domain of the probability distributions considered through a
            separate spatial metric. Combined with kinetically informed
            coordinate transforms such as TICA\cite{perez2013identification} and
            variants,\cite{noe2015kinetic,noe2016commute} it may be possible to
            parameterize models to have stationary
            distributions that are
            kinetically close to one another.

        \item

            The effect of an incomplete $\fpairspace$ should be
            investigated. In this case the presented divergence based
            interpretation is not trivially accurate.\cite{liu2018inductive}
            Understanding of how imperfect classifiers affect the
            parameterization of approximate models may have large implications
            on the optimization of complex multicomponent systems; overly
            expressive $\fpairspace$ will likely impede model parameterization
            as more sampling of the CG and FG system may be required.

    \end{enumerate}

\section{Concluding Remarks}\label{conc}

    In this article we discussed a new class of methods for the systematic
    bottom-up parameterization of a CG model.  In doing so we illustrated
    concrete connections between CG models and algorithms such as generative
    adversarial networks. Utilizing these connections we both decoupled the
    resolution at which we critique our CG model from the CG potential itself
    and enabled the use of a variety of novel measures of quality for CG model
    parameterization. We provided a proof of concept implementation and several
    numerical examples.  Additionally, we illustrated precise connections to
    several previous methods for CG model parameterization.  Finally, we noted
    multiple future branches of studies that can now be pursued. Together, these
    results open a new conceptual basis for future systematic CG
    parameterization strategies.

\section*{Acknowledgments}

    This material is based upon work supported by the
    National Science Foundation (NSF Grant CHE-1465248).
    This research was conducted with Government
    support under and awarded by DoD, Air Force Office of
    Scientific Research, National Defense Science and
    Engineering Graduate (NDSEG) Fellowship, 32 CFR 168a.
    Simulations were performed using the resources provided
    by the University of Chicago Research Computing Center
    (RCC).

\cleardoublepage

\appendix
\section{Envelope Theorem}\label{envthm}
    
    The envelope theorem is used to justify optimizing $\cgParam$ using only the
    partials calculated  holding the optimal
    observable constant.  A general statement of the envelope theorem is given
    by theorem 1 in \citet{milgrom2002envelope}, which also notes that the
    theorem applies to directional derivatives in a normed vector space.

    Let $X$ denote the choice set and let the relevant parameter be $t\in [0,1]$.
    Let $f:X\times[0,1] \to \mathbf{R}$ denote the parameterized objective
    function. The value function $V$ and the optimal choice correspondence
    $X^*$ are given by:
    \begin{equation}
        V(t) := \sup_{x\in X} f(x,t)
    \end{equation}
    \begin{equation}
        X^*(t) := \{x \in X: f(x,t) = V(t)\}
    \end{equation}
    Take $t \in (0,1)$ and $x^* \in X^*(t)$, and suppose that $\partial_t
    f(x^*,t)$ exists. If $V$ is differentiable at $t$, then $V'(t) =
    \partial_t f(x^*,t)$.

    This result puts no constraint on $X$, which corresponds to $\fpairspace$ in
    the current work, except that its maximal member have a derivative at that
    point. Additionally, as noted in \citet{milgrom2002envelope},
    this results is only useful if $V$ is known to be differentiable. This is
    compatible with our $f$-divergence variational statement when considered in
    the context of a complete $\fpairspace$ and population averages, but must in
    general be confirmed for each choice of $\fpairspace$. In situations where a
    closed form expression corresponding to the maximum is not known,
    constraints may be put on each member of $\fpairspace$ to ensure
    applicability. Suitable constraints may be found in the remainder of
    \citet{milgrom2002envelope}.

\section{Momentum Consistency}

    The described approach to achieve momentum consistency requires that we put
    more specific constraints on $\vmap$.  This is needed due to our minimal
    strategy for providing sufficiency conditions for consistency: primarily, we
    utilize arguments in previous work to provide sufficient constraints.
    The resulting conditions given suffice for the case of virtual particles
    that are simply dropped from the system by $\vmap$.  Generalizations to
    linear mappings that share particles between sites can additionally be inferred.
    First we discuss the approach of previous work on momentum consistency as is
    relevant to our work, and then concisely give a route to momentum
    consistency.

    \subsection{MS-CG}

        Generally, we will here assume that $\rmap$ satisfies specific
        properties. Once $\rmap$ is defined, we construct an appropriate
        $\pmap$.  First, $\rmap$ must be expressible in the following linear
        form, where $\rmap_I$ denotes the $I$th particle entry of the output of
        $\rmap$, $i$ iterates over the particles contribute to site $I$,
        and $c$ denotes positive constants.
        \begin{equation}\label{rmap_def}
            \rmap_I(r^\fgnp) := \sum_{i=1}^{n^\map_I} c^{\rmap}_{Ii} r_i
        \end{equation}
        As in MS-CG,\cite{noid2008multiscale} we impose translational consistency.
        \begin{equation}
            \sum_{i=1}^{n^\map_I} c^{\rmap}_{Ii} = 1
        \end{equation}
        From this we allow $\rmap$ to imply $\pmap$ up to the factor of the
        CG masses $\{M_I\}_I$ as stated in MS-CG.
        \begin{equation}
            \pmap_I(\mathbf{\fgmom}^\fgnp) := 
            M_I \sum_{i=1}^{n^\map_I} 
            \frac{c_{Ii} \fgmom_i}{m_i}
        \end{equation}
        As before, this type of map transforms global consistency into
        constituent momentum and position space components,
        i.e.,
        \begin{widetext}
            \begin{equation}
                \cgDensity(\pos^{3\cgnp},\mom^{3\cgnp}) = 
                \int \md\fgpos^{3\fgnp} \int \md \fgmom^{3\fgnp}
                \preDensity(\fgpos^{3\fgnp},\fgmom^{3\fgnp})
                \delta(\rmap(\fgpos^{3\fgnp}) - \pos^{3\cgnp})
                \delta(\pmap(\fgmom^{3\fgnp}) - \mom^{3\cgnp})
                = \cgDensityPos(\pos^{3\cgnp})\cgDensityMom(\mom^{3\cgnp})
            \end{equation}
        \end{widetext}
        where the vector valued delta functions are understood to be products of
        scalar delta functions. 
        If $\map$ does not associate any individual atoms to more than a
        single CG site, then
        \begin{eqnarray}
            \nonumber
            \exp\left(- \beta \sum_{I=1}^N \frac{\mom_I^2}{2 M_I} \right)
            &\propto&
            \int \md\fgmom^{3\fgnp} \exp\left( - \beta \sum_{i=1}^\fgnp 
            \frac{\fgmom_i^2}{2 m_i}
            \right)
            \\* 
            &&
            \times
            \delta(\pmap(\fgmom^{3\fgnp}) - \mom^{3\cgnp})
        \end{eqnarray}
        with
        \begin{equation}
            {M^\map_I}^{-1} := \sum_{i\in \mathcal{I}_I}
            \frac{{c^\map_{Ii}}^2}{m_i}
        \end{equation}
        We will additionally assume
        that analogous constraints are put on $\rvmap$ when considering momentum
        consistency below.

    \subsection{Momentum Consistency}

        Using these points we now move forward directly discussing momentum
        consistency. As stated previously, by constraining $\vmap$ and $\map$
        as above, and assuming the underlying systems are characterized by
        separable probability densities, we find
        \begin{eqnarray}
            \cgDensity(\pos^{3\cgnp},\mom^{3\cgnp}) 
            &= 
            &\cgDensityPos(\pos^{3\cgnp})\cgDensityMom(\mom^{3\cgnp})
            \\*
            \refDensity(\pos^{3\cgnp},\mom^{3\cgnp}) 
            &= 
            &\refDensityPos(\pos^{3\cgnp})\refDensityMom(\mom^{3\cgnp})
        \end{eqnarray}
        As a result, we split up our consistency statement (omitting arguments for clarity)
        \begin{equation}
            \left(\cgDensityPos
            = \refDensityPos
            \land
            \cgDensityMom
            = \refDensityMom
            \right)
            \implies
            \cgDensity = \refDensity
        \end{equation}
        Configurational consistency is handled via divergence matching
        as described in the main article; we
        here consider momentum consistency algebraically.
        \begin{widetext}
            \begin{equation}
                \cgDensityMom= \refDensityMom
                \iff
                \int \md\premom^{3\prenp}
                \exp\left( - \beta \sum_{i=1}^{\prenp} 
                \frac{\premom_i^2}{2 \premass_i}
                \right)
                \delta(\pvmap (\premom^{3\prenp}) - \mom^{3\cgnp})
                \propto
                \int \md\fgmom^{3\fgnp} 
                \exp\left( - \beta\sum_{i=1}^{\fgnp}
                \frac{\fgmom_i^2}{2 m_i}
                \right)
                \delta(\pmap(\fgmom^{3\fgnp}) - \mom^{3\cgnp}).
            \end{equation}
        \end{widetext}
        We substitute these using two sets of properly designed CG masses, each
        set implied by a mapping operator and the masses at resolution it maps
        \begin{eqnarray}
            \exp\left(- \beta \sum_{I=1}^N \frac{\mom_I^2}{2 M^\vmap_I} \right)
            &\propto&
            \exp\left(- \beta \sum_{I=1}^N \frac{\mom_I^2}{2 M^\map_I} \right)
            \\
            {M^\vmap_I}^{-1} &:=& \sum_{i\in \mathcal{I}^\vmap_I}
            \frac{{c^\vmap_{Ii}}^2}{\premass_i}
            \\
            {M^\map_I}^{-1} &:=& \sum_{i\in \mathcal{I}^\map_I}
            \frac{{c^\map_{Ii}}^2}{m_i}
            \label{momconsist}
        \end{eqnarray}
        The only solution is to set $M^\vmap_I = M^\map_I$ for each CG site
        $I$; in this case find a set of equations implying consistency.
        \begin{equation}\label{mass_crit}
            \left[
                0
                =
                 \sum_{i\in \mathcal{I}^\map_I}
                \frac{{c^\map_{Ii}}^2}{m_i}
                -
                \sum_{i\in \mathcal{I}^\vmap_I}
                \frac{{c^\vmap_{Ii}}^2}{\premass_i}
            \right]
            \forall \text{ CG sites }I
        \end{equation}
        Note that these equations are still subject to the aforementioned
        constraints (positivity, etc.).
        This provides a simple condition connecting our FG masses,
        pre-CG masses, $\map$, and $\vmap$, and allows one to check for momentum
        consistency.

        When considering a CG model with no pre-CG resolution the FG mapping
        $\map$ must associate each atom with at most a single CG site in order
        for the mapped momentum distribution to factorize with respect to each
        CG site. This is required for momentum consistency if the CG model is
        simulated using traditional molecular dynamics software as the momentum
        distribution produced by traditional molecular dynamics is necessarily
        factorizable. This same constraint to $\map$ and $\vmap$ is assumed in
        the preceding analysis, but this is not generally required for ARCG
        models as nonfactorizable momentum distributions may be produced by both
        $\map$ and $\vmap$. However, the analysis provided to illustrate momentum
        consistency is based on a generalizable strategy: previous approaches to
        momentum consistency which produced a closed form expression for a
        function proportional to the Boltzmann density of the mapped atomistic
        distribution may be extended to the current setting by simply
        calculating the density implied by both $\map$ and $\vmap$ and setting
        them to be equivalent.  In this way, more sophisticated approaches such
        as the one in \citet{han2018mesoscopic} may be applied analogously to
        approach more complex mapping operators.

\section{Loss derivations}\label{loss_derivation}

    The basis of the duality central to $f$-divergences is translated from
    theorem
    9 in \citet{reid2011}. The equations relating loss functions $l$ from the
    combined loss $\bloss$ may be confirmed via algebra after the two following
    identities are noted, both of which may be found in \citet{reid2011}.
    \begin{equation}
        \left. \frac{\partial\bloss}{\partial x}\right\rvert_h
        =
        l_\text{ref}(h) - l_\text{mod}(h)
    \end{equation}
    \begin{equation}
        \bloss(h)
        =
        (1-h) l_\text{mod}(h) + h l_\text{ref}(h)
    \end{equation}

    The terms needed to define $l_\text{ref}$ and $l_\text{mod}$ are given as
    follows. First, note
    that the function generating the appropriate relative entropy is $x \log x$
    (not $\log x$). From this we find (only in the case of relative entropy)
    \begin{equation}
        \bloss(h) = -2 x \log \frac{x}{1-x}
    \end{equation}
    and
    \begin{equation}
        \left. \frac{\partial\bloss}{\partial x}\right\rvert_h
        =
        -2 \left( \log \frac{h}{1-h} + \frac{1}{1-h}\right).
    \end{equation}
    Through substitution we then arrive at Eq. \eqref{re_losses}. A similar
    procedure may be used to emulate other $f$-divergences.

\section{Numerical Simulation Details}\label{num_details}

    This appendix contains details of the molecular potentials used, 
    the features used as input to the variational estimator, and the noise used
    to smooth the optimization.

    \subsection{Spline potentials}

    The $b$-splines describing the potential used to approximate water used
    knots that were not spaced evenly. Instead, various uniform regions of high
    and low knot density were used. This was due to computational constraints on
    the current implementation used to numerically optimize the potentials, not
    limitations of the methodology itself. It was found that a high knot density
    was needed to capture the inner well of the potential. The knots used for
    the water potential were 0., 0.417, 0.833, 1.25, 1.67, 2.08, 2.5, 2.6, 2.7,
    2.8, 2.9, 3.0, 3.1, 3.2, 3.3, 3.4, 3.5, 3.6, 3.7, 3.8, 3.9, 4.0, 4.1, 4.2,
    4.3, 4.4, 4.5, 4.6, 4.7, 4.8, 4.9, 5.0, 5.4, 5.8, 6.2, 6.6, and 7.0
    Angstroms. This corresponds to a higher density of knots near the inner
    well.  In contrast, the methanol potential instead used uniform knots spaced
    from 0 to 10 Angstroms. 

    \subsection{Variational features}

    This subsection describes the features used as input to the variational
    estimator. The single component LJ fluid and the integrated bonded particle
    used relatively simple feature sets, while
    the examples of the integrated binary LJ system, the approximated methanol
    system, and the approximated water system used a more complex 
    feature set as input the variational estimator. We here define the classes
    of features used, and then describe the set used for each of those examples.
    These features are calculated on each frame to produce the input for the
    variational estimator. 

    The first class of features is defined as
    the frame-wise average of a power of the distances between all the particles
    in the system.
    \begin{equation}
        H_{\mathrm{moment}} (\pos^{3\cgnp},n) = \frac{1}{n_p} \sum_{i>j} r_{ij}^n
    \end{equation}
    where $n_p$ is the number of pairs in the system.

    The second class characterizes the average local density of each frame. The
    local environment is characterized by passing the softened number of
    neighbors within a certain cutoff through a hyperbolic tangent function.
    Note that an offset and scaling factor is applied to this local density
    before the hyperbolic tangent is applied.

    \begin{equation}
        H_{\mathrm{density}} (\pos^{3\cgnp},r_{\mathrm{cut}},a,b) = 
        \frac{1}{n}
        \sum_{i} f \left(
            \frac{  \sum_{j \neq i} -g(r_{ij} - r_{\mathrm{cut}})-a}{b}
        \right)
    \end{equation}
    where $f$ is the hyperbolic tangent, $g$ is the logistic sigmoid, and $n$ is
    the number of particles in the system.

    The third class of features is given by calculating an RDF at each frame,
    i.e. given a radial bin it returns the number of particle pairs whose
    separating distance is in that bin.
    \begin{equation}
        H_{\mathrm{RDF}} (\pos^{3\cgnp},B) := 
        \frac{1}{n_p}\sum_{i>j} \mathbf{1}[r_{ij}\in B]
    \end{equation}

    The single component LJ system used $H_{\text{moment}}(\cdot,-6)$ and
    $H_{\text{moment}}(\cdot,-12)$. This set of features is sufficient to create
    a complete $\fpairspace$ as we are able to write the potential of both the
    reference and models systems as a function of it.  The virtual bonded
    particle only used the distance between the two real particles as input;
    this is can be seen to be sufficient by considering the rotational and
    translational symmetry present in the system.  The integrated binary LJ
    system and the approximated methanol system used the same set of features:
    this was composed of features from the 3 classes described above. The
    $H_{\mathrm{moment}}$ features were parameterized with 2, 4, 6, and 12. The
    parameterization of the $H_{\mathrm{density}}$ features is given in table
    \ref{table:feat_params}. The $H_{\mathrm{RDF}}$ features were parameterized
    with 50 equally spaced bins from 2.5 $\AA$ to 10 $\AA$. The featurization
    used the water example was identical except for the RDF features: in this
    case, they were parameterized 2 $\AA$ to 12 $\AA$ with 100 bins. The
    extended radial features were due to the higher resolution knot density.

    \begin{table}[h]
        \centering
        \renewcommand*{\arraystretch}{1.3}
        \begin{tabular}{| c | c | c | c |}
            \hline
            $r_{\mathrm{cut}}/\angstrom$ & $a$ & 
            $b$ \\ \hline
             4 & 0 & 2  \\ \hline
             4 & 1 & 2  \\ \hline
             7 & 7 & 2 \\ \hline
             7 & 9 & 2  \\ \hline
             10 & 9 & 2  \\ \hline
             10 & 11 & 2  \\ \hline
        \end{tabular}
        \caption{Parameters used for the local density feature functions.}
        \label{table:feat_params}
    \end{table}

    The neural network architectures used were simple feed-forward networks.
    Not including the input and output layers, the LJ example used to layers of
    5 nodes, the virtual bonded site example used 3 layers of 10 nodes, and the
    binary LJ system used 4 layers of 15 nodes. The architecture did not have a
    noticeable effect as long as at least two layers were present.

    \subsection{Noise}

    Noise was added to improve convergence of a variety of the numerical
    examples in this paper (all except the case of the virtual bonded particle).
    This is helpful in the cases examined when the distributions being optimized
    are highly dissimilar. The procedure used to apply noise is summarized as
    follows.  First, a data set composed of the combined samples from both the
    reference and model trajectories are whitened to have a mean of zero and a
    standard deviation of one in each dimension.  Gaussian noise was applied of
    a specified variance with a mean of zero was applied to each dimension. This
    variance noise was geometrically decayed when the reported accuracy of the
    classifier (produced by optimization of the variational statement) was below
    a set threshold for a set number of iterations. The decay factor was set to
    0.95-0.97 for the examples presented. Additional details be found in the
    tests presented in the public code base.

\bibliography{technical.bib}



\end{document}